\documentclass[a4paper,oneside,onecolumn,thmsb,12pt]{article}
\usepackage{sw20lart}

\input tcilatex
\QQQ{Language}{
American English
}

\begin{document}

\title{Coupled Dyson-Schwinger Equations and Effects of Self-Consistency}
\author{S. S. Wu, H. X. Zhang, Y. J. Yao \\
Center for Theoretical Physics and Department of Physics\\
Jilin University, Changchun 130023\\
People's Republic of China}
\date{}
\maketitle

\begin{abstract}
Using the $\sigma -\omega $ model as an effective tool, the effects of
self-consistency are studied in some detail. A coupled set of
Dyson-Schwinger equations for the renormalized baryon and meson propagators
in the $\sigma -\omega $ model is solved self-consistently according to the
dressed Hartree-Fock scheme, where the hadron propagators in both the baryon
and meson self-energies are required to also satisfy this coupled set of
equations. It is found that the self-consistency affects the baryon spectral
function noticeably, if only the interaction with $\sigma $ mesons is
considered. However, there is a cancellation between the effects due to the $%
\sigma $ and $\omega $ mesons and the additional contribution of $\omega $
mesons makes the above effect insignificant. In both the $\sigma $ and $%
\sigma -\omega $ cases the effects of self-consistency on meson spectral
function are perceptible, but they can nevertheless be taken account of
without a self-consistent calculation. Our study indicates that to include
the meson propagators in the self-consistency requirement is unnecessary and
one can stop at an early step of an iteration procedure to obtain a good
approximation to the fully self-consistent results of all the hadron
propagators in the model, if an appropriate initial input is chosen. Vertex
corrections and their effects on ghost poles are also studied.

\begin{enumerate}
\item[PACS]  :21. 60 J$_z$, 21. 65. +f, 11. 10. Gh.

\item[Keywords]  : Dyson-Schwinger Equations, effects of self-consistency.
\end{enumerate}
\end{abstract}

\section{. Introduction}

For a strong coupling theory it is important to go beyond the perturbation
calculation. The Dyson-Schwinger (DS) equation provides one of the effective
means to achieve such a goal. As is wellknown, the requirement of
self-consistency can make the infinite series of Feynman diagrams contained
in the solution to the equation nested, with each nest again containing the
same infinite series of nested diagrams. However, the equation will then
become nonlinear and not easy to solve. There is always the question what
new effects on the solution may be produced by such a nesting and whether
the latter is always important. In the zero-density case Brown, Puff and
Wilets [1] as well as Bielajew and Serot [2] have solved the DS equation for
the renormalized baryon propagators self-consistently according to the
dressed Hartree-Fock (HF) scheme. As is wellknown, the $\sigma -\omega $
model has achieved considerable success in the description of many bulk and
single particle properties of nuclei [3]. Thus it is worthwhile to use it as
a tool to study the effect of self-consistency in some detail. For
simplicity we shall only consider the case of zero-density, where the
tadpole self-energies need not be taken into account. The different
self-consistent (dressed) HF schemes are illustrated in Fig. 1. By
''dressed'' it is meant that in the baryon and meson self-energies the
hadron propagators are not all bare. If one requires self-consistency in the
baryon propagators, i.e. $\overline{G}=G$ in Fig. 1a, while sets $\overline{%
\Delta }=\Delta ^0$ and $\overline{D}=D^0$, it is referred to as scheme BP.
In this case the DS equation represented by Fig. 1a is already closed. Since 
$\overline{G}=G$, it is easily seen by iteration that the self-energy is now
nested and the nesting repeats again and again. Clearly one may also require
self-consistency in all the hadron propagators contained in the
self-energies. This implies that one should have $\overline{G}=G$, $%
\overline{\Delta }=\Delta $ and $\overline{D}=D$ and it will be referred to
as the fully self-consistent (FSC) scheme. In order to solve for $G$, $%
\Delta $ and $D$ we must now consider a coupled set of DS equations, since
the propagators are mutually correlated as shown in Fig. 1. Such a set of
equations has been solved recently by Bracco, Eiras, Krein and Wilets (BEKW)
[4] for a $\pi -\omega $ model. Their calculation shows that the effect of
self-consistency on the spectral properties of the hadron propagators is not
significant and it also can not make the ghost poles disappear. Following
previous works by Milana [5], Allendes and Serot [6] and Krein, Nielsen,
Puff and Wilets [7], BEKW further found that the ghost poles in all the
hadron propagators considered in the model can be eliminated if vertex
corrections are properly taken into account. Recently, by means of a simple
model we have shown [8] that the self-consistency in scheme BP diminishes
the continuum part of the spectral representation for the baryon propagator
and for the calculation of the latter, there is a simpler scheme which is a
good approximation to scheme BP. We would like to ascertain whether the
above results hold more generally. We shall study the coupled set of DS
equations according to the FSC scheme. The effects of self-consistency will
be investigated in some detail. It is found that they affect the baryon
spectral function noticeably, in case only the $\sigma $ meson is
considered. However, there is almost no difference between the results
obtained from scheme BP and FSC. It is interesting to find that the
additional contribution of $\omega $ mesons makes the above effect
insignificant. This indicates that there is a cancellation between the
contributions from the $\sigma $ and $\omega $ mesons. In both the $\sigma $
and $\sigma -\omega $ cases the effects of self-consistency on meson
spectral functions are perceptible, but they can be taken account of without
a self-consistent calculation. Our calculation shows that to include the
meson propagators in the self-consistent requirement is unnecessary and one
can stop at an early step of an iteration procedure to obtain a good
approximation to the FSC scheme for the calculation of all the hadron
propagators in the model. In analogy to Ref. [4], we have also obtained that
neither scheme BP nor scheme FSC can eliminate the ghost poles. Since the
question of ghost poles is of vital theoretical importance, following Refs.
[4, 6, 7, 9], we have further calculated the contributions of vertex
corrections and studied their consequences. As both the $\omega $ and $%
\sigma $ mesons are taken into account, vertex functions are only expressed
by phenomenological form factors.

In Section 2 we shall consider the coupled set of DS equations for the
renormalized hadron propagators in the $\sigma -\omega $ model and its
reduction. The numerical results are given and discussed in Section 3. The
vertex corrections and their consequences are studied in Section 4.
Concluding remarks are presented in Section 5.

\section{. The model and coupled Dyson-Schwinger equations}

The Lagrangian density for the $\sigma -\omega $ model [3] is given by 
\begin{equation}
\begin{array}{c}
L=-\overline{\psi }(\gamma _\mu \partial _\mu +M)\psi -\frac 12(\partial
_\mu \phi \partial _\mu \phi +m_s^2\phi ^2)-\frac 14F_{\mu \nu }F_{\mu \nu }
\\ 
-\frac 12m_v^2A_\mu A_\mu +g_s\overline{\psi }\psi \phi +ig_v\overline{\psi }%
\gamma _\mu \psi A_\mu +L_{CTC\text{ }}
\end{array}
,  \tag{1}
\end{equation}
where $F_{\mu \nu }=\partial _\mu A_\nu -\partial _\nu A_\mu $, $\partial
_\mu =\frac \partial {\partial x_\mu }$, $x_\mu =(\vec x,ix_0)$, $x^2=x_\mu
x_\mu =\vec x^2-x_0^2$ with $x_0\equiv t$ and CTC means the counterterm
correction introduced for the purpose of renormalization. The baryon, $%
\sigma $ and $\omega $ meson propagators are designated as follows: 
\begin{equation}
G_{\alpha \beta }(x=x_1-x_2)=\langle T[\psi _\alpha (x_1)\psi _\beta
(x_2)]\rangle =\dint \frac{d^4k}{(2\pi )^4}e^{ik_\rho x_\rho }G_{\alpha
\beta }(k),  \tag{2}
\end{equation}
\begin{equation}
\Delta (x)=\langle T[\phi (x_1)\phi (x_2)]\rangle =\dint \frac{d^4k}{(2\pi
)^4}e^{ik_\rho x_\rho }\Delta (k),  \tag{3}
\end{equation}
\begin{equation}
D_{\mu \nu }(x)=\langle T[A_\mu (x_1)A_\nu (x_2)]\rangle =\dint \frac{d^4k}{%
(2\pi )^4}e^{ik_\rho x_\rho }D_{\mu \nu }(k),  \tag{4}
\end{equation}
where $\langle 0\rangle \equiv \langle \overline{\Psi }_0|0|\overline{\Psi }%
_0\rangle $ and $|\overline{\Psi }_0\rangle $ denotes the exact ground
state. We note that our definition differs from the usual one by a factor of 
$i$. It is wellknown that the relevant DS equations in the dressed HF scheme
(Fig. 1) can be written in the following form:

(a) for baryon 
\begin{equation}
G(k)=G^0(k)+G^0(k)\Sigma (k)G(k)=-[\gamma _\mu k_\mu -iM+\Sigma (k)]^{-1}, 
\tag{5}
\end{equation}
\begin{equation}
\Sigma (k)=\Sigma _s(k)+\Sigma _v(k),  \tag{6a}
\end{equation}
\begin{equation}
\Sigma _s(k)=-g_s^2\dint \frac{d^\tau q}{(2\pi )^4}\overline{G}(q)\overline{%
\Delta }_s(k-q)\Gamma _s(k,q,k-q)+\Sigma _{CTC}^s(k),  \tag{6b}
\end{equation}
\begin{equation}
\Sigma _v(k)=g_v^2\dint \frac{d^\tau q}{(2\pi )^4}\gamma _\eta \overline{G}%
(q)\overline{D}_{\eta \lambda }(k-q)\Gamma _\lambda (k,q,k-q)+\Sigma
_{CTC}^v(k);  \tag{6c}
\end{equation}

(b) for $\sigma $-meson 
\begin{equation}
\Delta _s(k)=\Delta _s^0(k)+\Delta _s^0(k)\Pi _s(k)\Delta
_s(k)=-i[k^2+m_s^2+i\Pi _s(k)-i\epsilon ]^{-1},  \tag{7}
\end{equation}
\begin{equation}
\Pi _s(k)=g_s^2\dint \frac{d^\tau q}{(2\pi )^4}Tr[\overline{G}(k+q)\Gamma
_s(k+q,q,k)\overline{G}(q)]+\Pi _{CTC}^s(k);  \tag{8}
\end{equation}

(c) for $\omega $-meson 
\begin{equation}
D_{\mu \nu }(k)=D_{\mu \nu }^0(k)+D_{\mu \eta }^0(k)\Pi _{\eta \lambda
}(k)D_{\lambda \nu }(k),  \tag{9}
\end{equation}
\begin{equation}
\widehat{\Pi }_{\eta \lambda }(k)=-g_v^2\dint \frac{d^\tau q}{(2\pi )^4}%
Tr[\gamma _\eta \overline{G}(k+q)\Gamma _\lambda (k+q,q,k)\overline{G}(q)]; 
\tag{10}
\end{equation}
where $\Gamma _s$ and $\Gamma _\lambda $ are the $\sigma $-baryon and $%
\omega $-baryon vertex functions, respectively (denoted by a heavy dot in
Fig. 1), a caret indicates that the quantity is not yet renormalized, $\tau
=4-\delta $ ($\delta \rightarrow 0^{+}$) and in Eq. (5) the Feynman
prescription $M\rightarrow M-i\epsilon $ is understood. In this section
vertex corrections will not be considered, thus $\Gamma _s(p,q,k)=1$ and $%
\Gamma _\lambda (p,q,k)=\gamma _\lambda $. Since the baryon current is
conserved, we have 
\begin{equation}
\widehat{\Pi }_{\mu \nu }(k)=(\delta _{\mu \nu }-\frac{k_\mu k_\nu }{k^2})%
\widehat{\Pi }_v(k),  \tag{11}
\end{equation}
which implies $\widehat{\Pi }_v(k)=\frac 13\dsum_\mu \widehat{\Pi }_{\mu \mu
}(k).$ From Eq. (11) one observes that the renormalized $\Pi _{\mu \nu }(k)$
can be obtained from 
\begin{equation}
\Pi _v(k)=\widehat{\Pi }_v(k)+\Pi _{CTC}^v(k).  \tag{12}
\end{equation}
To fix the renormalization counterterms we shall use the following
conditions 
\begin{equation}
\Sigma _\alpha \left( k\right) \mid _{\gamma _\mu k_\mu =iM_t}=0;\qquad
\frac \partial {\partial (\gamma _\mu k_\mu )}\Sigma _\alpha (k)|_{\gamma
_\mu k_\mu =iM_t}=0,  \tag{13a}
\end{equation}
\begin{equation}
\Pi _\alpha \left( k\right) \mid _{k^2=0}=0;\qquad \frac \partial {\partial
(k^2)}\Pi _\alpha (k)|_{k^2=0}=0,  \tag{13b}
\end{equation}
where $\alpha =s$ or $v$ and $M_{t\text{ }}$ denotes the true baryon mass.
If we set $\overline{G}=G$, $\overline{\Delta }_s=\Delta _s$ and $\overline{D%
}_{\eta \lambda }=D_{\eta \lambda }$ in Eqs. (6, 8 and 10), it is seen that
Eqs. (5-10) constitute a self-consistent closed set of DS equations for all
the hadron propagators considered in the model. It thus yields the FSC
scheme. We note that $D_{\eta \lambda }$ can be written as [10, 11] 
\begin{equation}
D_{\eta \lambda }(k)=(\delta _{\eta \lambda }-\frac{k_\eta k_\lambda }{k^2})%
\frac{-i}{k^2+m_v^2+i\Pi _v(k)-i\epsilon }-i\frac{k_\eta k_\lambda }{%
k^2(m_v^2+\delta m_v^2)},  \tag{14}
\end{equation}
where $\delta m_v^2$ is the mass counterterm for the $\omega $-meson. Since
the $\omega $-meson couples to the conserved baryon current, we may set $%
D_{\eta \lambda }=\delta _{\eta \lambda }\Delta _v$ in Eq. (6c), where
according to Eq. (14) $\Delta _v$ is given by 
\begin{equation}
\Delta _v(k)=\Delta _v^0(k)+\Delta _v^0(k)\Pi _v(k)\Delta
_v(k)=-i[k^2+m_v^2+i\Pi _v(k)-i\epsilon ]^{-1},  \tag{15}
\end{equation}
If we put $\overline{\Delta }_s=\Delta _s^0$ and $\overline{D}_{\eta \lambda
}=D_{\eta \lambda }^0$ and only require $\overline{G}=G$, Eqs. (5) and (6)
already build a closed self-consistent set, which has been called scheme BP.
It is known that the original HF approximation to $G(k)$ is given by 
\begin{equation}
G_\Sigma ^0(k)=-[\gamma _\mu k_\mu -iM+\Sigma (\vec k,E_k)]^{-1},  \tag{16}
\end{equation}
and the corresponding eigenvalue equation has the form 
\begin{equation}
\lbrack \gamma _\mu k_\mu -iM+\Sigma (k)]_{k_0=E_k}u(ks)=0.  \tag{17}
\end{equation}
If in addition to the approximations $\overline{\Delta }_s=\Delta _s^0$ and $%
\overline{D}_{\eta \lambda }=\delta _{\eta \lambda }\Delta _v^0$, we further
substitute $G_\Sigma ^0$ for $G$ in Eq. (6), we obtain 
\begin{equation}
\Sigma (k)=\dint \frac{d^\tau q}{(2\pi )^4}[-g_s^2G_\Sigma ^0(q)\Delta
_s^0(k-q)+g_v^2\gamma _\eta G_\Sigma ^0(q)\gamma _\eta \Delta _v^0(k-q)%
]+\Sigma _{CTC}(k),  \tag{18}
\end{equation}
Setting $k_0=E_k$ in Eq. (18), one obtains the self-consistent equation for
the HF potential $\Sigma (\vec k,E_k)\equiv \Sigma (k)|_{k_0=E_k}$ [12]. Let
the self-energy determined by Eq. (18) be indicated by a subscript $p$. We
note that $G_p(k)=-[\gamma _\mu k_\mu -iM+\Sigma _p(k)]^{-1}$ is different
from $G_\Sigma ^0(k)$, because $\Sigma _p(k)$ is defined for all $\vec k$
and $k_0$, though $\Sigma _p(k)|_{k_0=E_k}=\Sigma (\vec k,E_k)$. It has been
shown in Ref. [8] that $G_p(k)$ is a better approximation to $G(k)$
determined by scheme BP than $G_\Sigma ^0(k)$ and such an approximation has
been denoted as scheme P. Using $G_p$ we shall show that there is an
extended potential scheme which yields a good approximation to the FSC
scheme for the calculation of all the hadron propagators in the model. The
four-dimensional integrals involved in Eqs. (6, 8, 10) can be calculated by
means of the spectral representation [1]. Following Ref. [2], we shall apply
Feynman's integral parameterization to simplify the integration and use the
standard method employed in the dimensional regularization to evaluate the
relevant integrals. As is wellknown [13, 11], a properly normalized spectral
representation for the baryon propagator may be expressed in the form 
\begin{equation}
\widetilde{G}(k)=-Z_2\frac{\gamma _\mu k_\mu +iM_t}{k^2+M_t^2-i\epsilon }%
-\dint_{m_1^2}^\infty dm^2\frac{\gamma _\mu k_\mu \alpha
(-m^2,Z_2)+iM_t\beta (-m^2,Z_2)}{k^2+m^2-i\epsilon },  \tag{19a}
\end{equation}
where $m_1=M_t+m_s$ and $Z_2$ satisfies 
\begin{equation}
Z_2+\int_{m_1^2}^\infty dm^2\alpha (-m^2,Z_2)=1;  \tag{19b}
\end{equation}
whereas for the meson propagators we have 
\begin{equation}
\widetilde{\Delta }_\lambda (k)=Z_\lambda \frac{-i}{k^2+\widetilde{m}%
_\lambda ^2-i\epsilon }-i\dint_{th_\lambda }^\infty dm^2\frac{\rho _\lambda
(-m^2,Z_\lambda )}{k^2+m^2-i\epsilon },  \tag{20a}
\end{equation}
where $\widetilde{m}_\lambda $ is the true meson mass, $\lambda =s$ or $v$, $%
th_\lambda $ means the threshold where the continuum starts and 
\begin{equation}
Z_\lambda +\int_{th_\lambda }^\infty dm^2\rho _\lambda (-m^2,Z_\lambda )=1. 
\tag{20b}
\end{equation}
Since in Eq. (1) we have only considered the linear meson interaction and no
meson self-interactions are included, from Eqs. (8) and (10) it is easily
seen that $th_\lambda =(2M_t)^2$. Comparing Eq. (19a) with Eq. (5) and Eq.
(20a) with Eq. (7) as well as (15), we have $\widetilde{G}(k)=ZG(k)$ and $%
\widetilde{\Delta }_\lambda (k)=Y_\lambda \Delta _\lambda (k)$, where $Z$
and $Y_\lambda $ are proportional factors. It is not difficult to see that $%
Z_2=ZZ_t$ where ($-Z_t$) is the residue of $G(k)$ at the pole $\gamma _\mu
k_\mu =iM_t$, while under the on-shell renormalization condition (13a) one
has $Z_t=1$ and $M=M_t$ [8]. Similarly, for the meson propagators $Z_\lambda
=Y_\lambda R_\lambda $, where ($-iR_\lambda $) is the residue of $\Delta
_\lambda (k)$ at the pole $k^2=-\widetilde{m}_\lambda ^2$. Denote the
denominator of $\Delta _\lambda (k)$ by $D_\lambda (k^2)$ ($\Pi _\lambda $
is a function of only $k^2$). The pole ($-\widetilde{m}_\lambda ^2$) should
clearly satisfy $D_\lambda (-\widetilde{m}_\lambda ^2)=0$. From $D_\lambda
(k^2)=D_\lambda (-\widetilde{m}_\lambda ^2)+(k^2+\widetilde{m}_\lambda ^2)[%
\frac{dD_\lambda }{dk^2}]_{k^2=-\widetilde{m}_\lambda ^2}+\cdots $, one
finds immediately 
\begin{equation}
R_\lambda =[1+i(\partial \Pi _\lambda /\partial k^2)_{k^2=-\widetilde{m}%
_\lambda ^2}]^{-1}.  \tag{21}
\end{equation}
It is known [2] that in the zero-density case one may write 
\begin{equation}
\Sigma (k)=\gamma _\mu k_\mu a(k^2)-iMb(k^2).  \tag{22}
\end{equation}
Denote $a=a_s+a_v$ and $b=b_s+b_v$. Substituting 
\begin{equation}
\overline{G}(k)=Z^{-1}\widetilde{G}(k)=-\dint_0^\infty d\sigma ^2\frac{%
\gamma _\mu k_\mu f_\alpha (-\sigma ^2)+iM_tf_\beta (-\sigma ^2)}{k^2+\sigma
^2-i\epsilon },  \tag{23a}
\end{equation}
\begin{equation}
f_\gamma (-\sigma ^2)=\delta (\sigma ^2-M_t^2)+\theta (\sigma
^2-m_1^2)\gamma (-\sigma ^2),\qquad (\gamma =\alpha \;\text{or}\;\beta ) 
\tag{23b}
\end{equation}
\begin{equation}
\overline{\Delta }_\lambda (k)=Y_\lambda ^{-1}\widetilde{\Delta }_\lambda
(k)=-i\dint_0^\infty dm^2\frac{h_\lambda (-m^2)}{k^2+m^2-i\epsilon }, 
\tag{23c}
\end{equation}
\begin{equation}
h_\lambda (-m^2)=R_\lambda \delta (m^2-\widetilde{m}_\lambda ^2)+\theta
(m^2-th_\lambda )\rho _\lambda (-m^2),\qquad (\lambda =s\;\text{or}\;v) 
\tag{23d}
\end{equation}
into Eq. (6) and following the procedure suggested in Ref. [2], one easily
finds 
\begin{equation}
a_s(k^2)=\frac{g_s^2}{16\pi ^2}\diint_0^\infty d\sigma
^2dm^2\dint_0^1dxf_\alpha (-\sigma ^2)h_s(-m^2)x\ln \frac{K^2(-M_t^2)}{%
K^2(k^2)}+c_{s,}  \tag{24a}
\end{equation}
\begin{equation}
b_s(k^2)=\frac{g_s^2}{16\pi ^2}\diint_0^\infty d\sigma
^2dm^2\dint_0^1dxf_\beta (-\sigma ^2)h_s(-m^2)\ln \frac{K^2(k^2)}{K^2(-M_t^2)%
}+c_{s,}  \tag{24b}
\end{equation}
\begin{equation}
c_s=-\frac{g_s^2}{8\pi ^2}M_t^2\diint_0^\infty d\sigma ^2dm^2\dint_0^1dx%
\frac{x(1-x)[xf_\alpha (-\sigma ^2)+f_\beta (-\sigma ^2)]h_s(-m^2)}{%
K^2(-M_t^2)},  \tag{24c}
\end{equation}
\begin{equation}
K^2(k^2)\equiv K^2(x,\sigma ^2,m^2,k^2)=x(1-x)k^2+(1-x)\sigma ^2+xm^2; 
\tag{24d}
\end{equation}
\begin{equation}
a_v(k^2)=\frac{g_v^2}{8\pi ^2}\diint_0^\infty d\sigma
^2dm^2\dint_0^1dxf_\alpha (-\sigma ^2)h_v(-m^2)x\ln \frac{K^2(-M_t^2)}{%
K^2(k^2)}+c_{v,}  \tag{25a}
\end{equation}
\begin{equation}
b_v(k^2)=\frac{g_v^2}{4\pi ^2}\diint_0^\infty d\sigma
^2dm^2\dint_0^1dxf_\beta (-\sigma ^2)h_v(-m^2)\ln \frac{K^2(-M_t^2)}{K^2(k^2)%
}+c_{v,}  \tag{25b}
\end{equation}
\begin{equation}
c_v=\frac{g_v^2}{4\pi ^2}M_t^2\diint_0^\infty d\sigma ^2dm^2\dint_0^1dx\frac{%
x(1-x)[2f_\beta (-\sigma ^2)-xf_\alpha (-\sigma ^2)]h_v(-m^2)}{K^2(-M_t^2)},
\tag{25c}
\end{equation}
where $\gamma (-\sigma ^2)=Z^{-1}\gamma (-\sigma ^2,Z_2)$, $\rho _\lambda
(-m^2)=Y_\lambda ^{-1}\rho _\lambda (-m^2,Z_\lambda )$ and 
\begin{equation}
\alpha (k^2)=\frac 1\pi Im\left[ \frac{1+a_s(k^2)+a_v(k^2)}{D(k^2)}\right] ,
\tag{26a}
\end{equation}
\begin{equation}
\beta (k^2)=\frac 1\pi Im\left[ \frac{1+b_s(k^2)+b_v(k^2)}{D(k^2)}\right] , 
\tag{26b}
\end{equation}
\begin{equation}
D(k^2)=(1+a_s(k^2)+a_v(k^2))^2k^2+(1+b_s(k^2)+b_v(k^2))^2M_t^2.  \tag{26c}
\end{equation}
It is seen that $h_\lambda (-m^2)$ is yet unknown and use must now be made
of Eqs. (7-10). Since Eqs. (8) and (10) can be worked out in the same way,
let us consider the latter. Substituting Eqs. (23a and b) into Eq. (10) and
using the Feynman formula $(AB)^{-1}=\smallint _0^\infty dx[xA+(1-x)B]^{-2}$%
, one gets 
\begin{equation}
\begin{array}{c}
\sum_\eta \widehat{\Pi }_{\eta \eta }(k)=8g_v^2\diint_0^\infty d\sigma
^2dm^2\dint_0^1dx\dint \frac{d^\tau Q}{(2\pi )^4} \\ 
\lbrack (Q^2-x(1-x)k^2)F_\alpha (\sigma ^2m^2)+2M_t^2F_\beta (\sigma
^2m^2)][Q^2+K^2(k^2)]^{-2}
\end{array}
,  \tag{27}
\end{equation}
where $F_\gamma (\sigma ^2m^2)=f_\gamma (-\sigma ^2)f_\gamma (-m^2)$. By
means of Eq. (13b) and the integration formula familiar in the dimensional
regularization we obtain that

\begin{equation}
\begin{array}{c}
\Pi _v(k^2)=\frac 13\sum_\eta [\widehat{\Pi }_{\eta \eta }(k^2)-\widehat{\Pi 
}_{\eta \eta }(0)-k^2(\partial \widehat{\Pi }_{\eta \eta }/\partial
k^2)_{k^2=0}] \\ 
=i\frac{g_v^2}{3\pi ^2}\diint_0^\infty d\sigma
^2dm^2\dint_0^1dx\{[(K^2(k^2)+\frac 12x(1-x)k^2)F_\alpha (\sigma ^2m^2)- \\ 
M_t^2F_\beta (\sigma ^2m^2)]\ln \frac{K^2(k^2)}{K^2(0)}-x(1-x)k^2[F_\alpha
(\sigma ^2m^2)-\frac{M_t^2}{K^2(0)}F_\beta (\sigma ^2m^2)]\},
\end{array}
\tag{28}
\end{equation}
where the counterterms are supplied by terms $\frac 12\delta m_v^2A_\mu
A_\mu $ and $\varsigma _v\frac 14F_{\mu \nu }F_{\mu \nu }$ contained in $%
L_{CTC}$ [see Eq. (1)]. Similarly we have 
\begin{equation}
\begin{array}{c}
\Pi _s(k^2)=i\frac{g_s^2}{4\pi ^2}\diint_0^\infty d\sigma
^2dm^2\dint_0^1dx\{[(2K^2(k^2)+x(1-x)k^2)F_\alpha (\sigma ^2m^2)+ \\ 
M_t^2F_\beta (\sigma ^2m^2)]\ln \frac{K^2(k^2)}{K^2(0)}-x(1-x)k^2[2F_\alpha
(\sigma ^2m^2)+\frac{M_t^2}{K^2(0)}F_\beta (\sigma ^2m^2)]\}.
\end{array}
\tag{29}
\end{equation}
Further, using Eqs. (7, 15, 20) and the formula $(x-i\epsilon )^{-1}=\frac
Px+i\pi \delta (x)$, we obtain 
\begin{equation}
\rho _\alpha (k^2)=\frac 1\pi Im[i\Delta _\alpha (k^2)],  \tag{30}
\end{equation}
for $k^2<-th_\lambda $. Eqs. (24-26) and (28-30) are the explicit
expressions for the closed set of renormalized DS equations used for the
calculation. We note that through the renormalization conditions (13a and b) 
$G(k)$ and $\Delta _\alpha (k)$ given by Eqs.(5, 7, 15) are normalized
differently from $\widetilde{G}(k)$ and $\widetilde{\Delta }_\lambda (k)$,
respectively. According to Eqs. (19b) and (20b) we shall define $I_b$ ($%
I_\lambda $) given below 
\begin{equation}
I_b\equiv Z\int_{m_1^2}^\infty dm^2\alpha (-m^2)=1-ZZ_t,  \tag{19c}
\end{equation}
\begin{equation}
I_\lambda \equiv Y_\lambda \int_{th_\lambda }^\infty dm^2\rho _\lambda
(-m^2)=1-Y_\lambda R_\lambda ,  \tag{20c}
\end{equation}
as the impurity factor of the single particle (sp) character of a baryon ($%
\lambda $-meson). Clearly, it just tells the relative importance of the
continuum part.

\section{. Numerical results}

Though the set of self-consistent DS equations for scheme FSC obtained in
Section 2 looks somewhat formidable, it can be solved rigorously and quite
easily by the method of iteration. Thus, it provides a useful example for
the study of self-consistency and diagram nesting. In the following the mass
values used are $M_t=4.7585$, $m_s=2.6353$ and $m_v=3.9680(fm^{-1})$. We
shall first consider the case of baryons interacting only with $\sigma $
mesons. Two values of the strength parameter $\overline{g}_s^2$($\equiv 
\frac{g_s^2}{16\pi ^2}$) are studied. Since the results of $\overline{g}%
_s^2=0.5263$ and $0.6517$ display the same qualitative behavior, only the
former will be represented graphically. From Figs. 2 and 3 one sees that the
results obtained from schemes BP and FSC are almost the same. Let us denote
a physical quantity $Q(k)$ by $Q(k;S)$ or simply $Q(S)$, if it is calculated
according to scheme $S$. Though one has set $\overline{\Delta }_s=\Delta
_s^0 $ in scheme BP, one may substitute the baryon propagator $G(k;BP)$
obtained from it into Eq. (8) and find $\rho _s(k^2;BP)$ from Eq. (30). It
is found that $\rho _s(k^2;BP)$ and $\rho _s(k^2;FSC)$ are very close to
each other. Thus, only the latter is shown in Fig. 4. The above results
indicate that to include the meson propagators in the self-consistency
requirement may not be necessary. Form Eqs. (5-10) it is seen that all the
self-energies will be known, if $\overline{G}$, $\overline{\Delta }_s$ and $%
\overline{D}_{\eta \lambda }$ are known. Thus different approximate schemes
are obtained, if different choices are made for $\overline{G}$, $\overline{%
\Delta }_s$ and $\overline{D}_{\eta \lambda }$. To assess the importance of
self-consistency and to ask whether one may stop at an earlier step of the
iteration process, we have shown in Fig. 3 results from five different
calculation schemes which are explained in Table 1. The first column gives
the name of each scheme, while the second and third explain how its $\Sigma
_s$ [$\Sigma _v$] and $\Pi _s$ [$\Pi _v$] are obtained from Eqs. (6b) [(6c)]
and (8) [(10)], respectively. For instance, $\Sigma _s(P)$ and $\Pi _s(P)$
in scheme P are obtained by setting $\overline{G}=G_\Sigma ^0$, $\overline{%
\Delta }_s=\Delta _s^0$ in Eq. (6b) and $\overline{G}=G_\Sigma ^0$ in Eq.
(8), $\Sigma _s(PEP)$ and $\Pi _s(PEP)$ in the partially extended P scheme
(PEP) by $\overline{G}=G(P)$, $\overline{\Delta }_s=\Delta _s^0$ in Eq. (6b)
and $\overline{G}=G(P) $ in Eq. (8), whereas to obtain $\Sigma _s(EP)$ and $%
\Pi _s(EP)$ for the extended P scheme (EP), one sets $\overline{G}=G(P)$, $%
\overline{\Delta }_s$ $=\Delta _s(P)$ in Eq. (6b) and $\overline{G}=G(P)$ in
Eq. (8), etc. The quantities in the brackets are for $\omega $-mesons in the 
$\sigma -\omega $ model. From Fig. 3 it is seen that the self-consistency
diminishes the continuum part of the spectral representation for the baryon
propagator [8], though the convergent process from ($\alpha (P),\beta (P)$)
to ($\alpha (FSC),\beta (FSC)$) is oscillatory. It is interesting to note
that the contribution of the meson self-consistency, though very small, adds
toward the same direction. Clearly we have 
\begin{equation}
\begin{array}{c}
\alpha (PEP)\simeq \alpha (EP),\quad \beta (PEP)\simeq \beta (EP) \\ 
\alpha (BP)\simeq \alpha (FSC),\quad \beta (BP)\simeq \beta (FSC)
\end{array}
,  \tag{31a}
\end{equation}
whereas Fig. 4 shows 
\begin{equation}
\rho _s(PEP)=\rho _s(EP)\simeq \rho _s(BP)\simeq \rho _s(FSC).  \tag{31b}
\end{equation}
In order to gain some quantitative insight into the above relations we have
calculated and listed the values $I_b$ ($I_\sigma $) of the baryon ($\sigma $%
-meson) sp impurity factor in Tables 2 and 3. In fact, the small values of $%
I_\sigma $ also explain why it is not important to require self-consistency
in the meson propagators. One observes that the difference between the
results of schemes PEP and EP is small and PEP seems a quite good
approximation to FSC with respect to all the hadron propagators considered.
To ascertain this we have further calculated 
\begin{equation}
\begin{array}{c}
F(\pm )\equiv \langle iG(k)\rangle _{\pm }=u_{\pm }^{+}(ks)iG(k)u_{\pm }(ks)
\\ 
\quad =\frac{1+a(k^2)}{D(k^2)}\left[ \frac{1+b(k^2)}{1+a(k^2)}\pm \frac{k_0}{%
E_k}\right] M_t
\end{array}
,  \tag{32}
\end{equation}
[8], where $+(-)$ refers to the eigenspinor of Eq. (17) with eigenvalue $%
+(-)E_k=[M_t^2+\vec k^2]^{\frac 12}$. In Fig. 5 we have only plotted $%
F_{r(i)}(FSC)=Re(Im)\langle iG(FSC)\rangle _{+}$, with $k_0=\sqrt{350}%
fm^{-1} $, since the numerical results of the three schemes $P$, $PEP$ and $%
FSC$ are nearly the same. Thus, if we use Eq. (32) as a criterion, we note
that even $G(P)$ is a good approximation. The poles of $G(k)$ and $\Delta
_\lambda (k)$ are given by the roots $k^2$ of $D(k^2)$ of Eq. (26c) with $%
a_v=b_v=0$ and 
\begin{equation}
D_\lambda (k^2)=k^2+m_\lambda ^2+i\Pi _\lambda (k^2),  \tag{33}
\end{equation}
respectively. The ghost poles found are listed in Table 4. They are given in
units of $fm^{-2}$.

Now let us consider the $\sigma -\omega $ model. Here we shall only present
the numerical results for $\overline{g}_s^2=0.5263$ and $\overline{g}%
_v^2\equiv \frac{g_v^2}{8\pi ^2}=1.3685$, as qualitatively they are quite
typical. In Figs. 6 and 7 we have plotted the numerical results for ($%
a_s,b_s $) and ($a_v,b_v$). It is seen that before $k^2\simeq -120fm^{-2}$
the effects of self-consistency is insignificant. Hereafter they seem to
become more important, but they are not large enough to cause any palpable
effects on $\alpha $ and $\beta $. The results obtained from the five
schemes nearly coincide with each other as shown in Fig. 8, where only two
are drawn explicitly, because the other four are not distinguishable in the
figure. Thus, in this case no self-consistent requirement is needed, just as
found in Ref. [4] for the $\pi -\omega $ model. Comparing Figs. 6 and 7 with
Fig. 2 one notes that $a_s(k^2)$ and $b_s(k^2)$ only change slightly due to
the presence of $\omega $ mesons. However, there is a quite large difference
between ($a_s,b_s$) and ($a,b$) where $a=a_s+a_v$ and $b=b_s+b_v$.This
explains why Fig. 8 differs from Fig. 3 so greatly. A comparison between
Figs. 3 and 8 suggests that there exists a cancellation between the effects
on self-consistency due to the $\sigma $ and $\omega $ mesons. This is
obviously related to the $\sigma $-baryon and $\omega $-baryon interactions
and thus also to the different characteristic features contributed to the
baryon-baryon interaction by the $\sigma $ and $\omega $ mesons
individually. It seems that these features are not only of importance to the
nuclear binding and saturation, but also to the formation of baryonic
excited states (see below). We expect the above relation between the
cancellation and the different characteristics of meson-baryon interactions
may hold generally, though the details of its regularity will depend on the
participating meson fields and there may even be enhancement if their
characteristics are of the same kind (which is actually defined by the
'enhancement'). If the regularity of cancellation can be extended to the
case of finite density, it will be very useful, because the analysis and the
renormalized calculation will be greatly simplified, when the effect of
self-consistency is negligible. However, this is still a problem to be
studied. To illustrate how the $\omega $ meson affects the disappearance of
the effect of self-consistency we have drawn a set of $\alpha $ and $\beta $
calculated for an intermediate $\overline{g}_v^2=0.3400$ in Fig. 9. We note
that owing to the presence of $\omega $ mesons the maximum of each baryon
spectral function becomes sharper and more distinct. Since $\alpha (k^2)$
relates directly to the probability of occurrence of an excited baryon
state, it indicates that the contribution of $\omega $ mesons enhances the
possibility of forming a resonant baryon state, though we do not expect that
the $\sigma -\omega $ model may explain the experimentally observed baryon
spectrum. From Fig. 10 it looks that a discernible effect of
self-consistency exists for mesons. However, we observe from the same figure
that schemes PEP and FSC yield almost the same result. This says that the
above effect can be taken account of by simply substituting $G(k;P)$ in Eqs.
(8) and (10) without considering a self-consistent calculation. BEKW [4]
found that for the $\pi -\omega $ model the self-consistency decreases $\rho
_\pi \left( k^2\right) $, while increases $\rho _\omega \left( k^2\right) $.
Owing to the difference in models, our calculation shows that the
self-consistency decreases both $\rho _s\left( k^2\right) $ and $\rho
_v\left( k^2\right) $. Further we note that the self-consistency and
additional contribution of $\omega $ mesons still cannot eliminate the ghost
poles, whose values found for the relevant hadron propagators are
respectively, for baryon: $149.4215\pm i96.8997$, for $\sigma $ meson: $%
274.4833$ and for $\omega $ meson: $178.5431(fm^{-2})$.

\section{. Effects of vertex corrections on ghost poles}

The principal aim of this section is to study the effects of vertex
corrections on ghost poles and their elimination. For our purpose we shall
only consider phenomenological form factors, though they still suffer the
not yet resolved problem of violating the Ward-Takahashi identity as
emphasized in Refs. [3, 7]. The vertices included in Eq. (1) are illustrated
in Fig. 11, where 1 and 2 denote baryons, while $\lambda $ a $\sigma $ or $%
\omega $ meson. Phenomenologically the vertex functions may be written
approximately as [14] 
\begin{equation}
\Gamma _s(p_1,p_2,p_s)=F_b(p_1)F_b(p_2)F_s(p_s),  \tag{34a}
\end{equation}
\begin{equation}
\Gamma _\mu (p_1,p_2,p_v)=\gamma _\mu F_b(p_1)F_b(p_2)F_v(p_v),  \tag{34b}
\end{equation}
where $F_b$ in $\Gamma _s$ and $\Gamma _\mu $ will be chosen the same.
Substituting Eq. (34) in Eqs. (6, 8, 10), we obtain 
\begin{equation}
\Sigma _s^f(k)=-g_s^2F_b(k)\dint \frac{d^\tau q}{(2\pi )^4}%
F_b(q)F_s(k-q)G(q)\Delta _s(k-q),  \tag{35a}
\end{equation}
\begin{equation}
\Sigma _v^f(k)=g_v^2F_b(k)\dint \frac{d^\tau q}{(2\pi )^4}%
F_b(q)F_v(k-q)\gamma _\mu G(q)D_{\mu \nu }(k-q)\gamma _\nu ,  \tag{35b}
\end{equation}
\begin{equation}
\Pi _s^f(k)=g_s^2F_s(k)\dint \frac{d^\tau q}{(2\pi )^4}%
F_b(k+q)F_b(q)Tr[G(k+q)G(q)],  \tag{36}
\end{equation}
\begin{equation}
\Pi _{\mu v}^f(k)=-g_v^2F_v(k)\dint \frac{d^\tau q}{(2\pi )^4}%
F_b(k+q)F_b(q)Tr[\gamma _\mu G(k+q)\gamma _\nu G(q)],  \tag{37}
\end{equation}
where the superscript $f$ indicates that form factors are inserted.

In the previous section it has been shown that all the calculated
renormalized hadron propagators possess ghost poles. Just as found in Refs.
[4, 6, 7, 9], our calculation also shows that the ghost poles can be
eliminated by a proper treatment of vertex corrections. In order to make the
change of the $k$-dependence of the self-energies caused by them appear in a
clearer form, we shall choose the form factors as simple as possible. We
shall set $F_b=1$ and 
\begin{equation}
F_\lambda (p)=\frac{\Lambda _\lambda ^2-m_\lambda ^2}{\Lambda _\lambda
^2+p^2-i\epsilon },\qquad \lambda =s\text{ or }v  \tag{38}
\end{equation}
where $\epsilon \rightarrow 0^{+}$ indicates the displacement of the poles.
From Eq. (35) it is immediately seen that $\Sigma _\lambda ^f$($\lambda =s$
or $v$) are made finite by this choice. However, according to Eqs. (36) and
(37) we have 
\begin{equation}
\Pi _s^f(k)=F_s(k)g_s^2\dint \frac{d^\tau q}{(2\pi )^4}Tr[G(k+q)G(q)]\equiv
F_s(k)\widehat{\Pi }_s(k),  \tag{39}
\end{equation}
\begin{equation}
\Pi _{\mu \nu }^f(k)=-F_v(k)g_v^2\dint \frac{d^\tau q}{(2\pi )^4}Tr[\gamma
_\mu G(k+q)\gamma _\nu G(q)]\equiv F_v(k)\widehat{\bar \Pi }_{\mu \nu }(k), 
\tag{40}
\end{equation}
where $\widehat{\Pi }_s$ ($\widehat{\bar \Pi }_{\mu \nu }$) is just the
integral in Eq. (8) (Eq. (10)) with $\Gamma _s=1$($\Gamma _\nu =\gamma _\nu $%
). Both of them are divergent. As no confusion will arise, the bar over $\Pi
_{\mu \nu }$ will hereafter be omitted. In order to investigate the effect
of the multiplicative factor $F_\lambda $ on the elimination of ghost poles,
we shall, just as Ref. [6], substitute the renormalized $\Pi _s$ and $\Pi
_{\mu \nu }$ for $\widehat{\Pi }_s$ and $\widehat{\Pi }_{\mu \nu }$,
respectively. The explicit expressions of $\Pi _{\mu \nu }$ and $\Pi _s$
have been given in Eqs. (11, 12, 28, 29).

Now let us consider the baryon self-energy. Substituting Eq. (38) into Eq.
(35) and using the method of integral parameterization, we get 
\begin{equation}
\begin{array}{c}
\Sigma _s^f(k)=-2ig_s^2\int \int d\sigma ^2dm^2(\Lambda _s^2-m_s^2)h_s(-m^2)
\\ 
\times \dint_0^1dx\dint_0^{1-x}dy\dint \frac{d^4Q}{(2\pi )^4}\frac{%
(x+y)\gamma _\mu k_\mu f_\alpha (-\sigma ^2)+iM_tf_\beta (-\sigma ^2)}{%
[Q^2+L_f^2-i\eta ]^3}
\end{array}
,  \tag{41a}
\end{equation}
\begin{equation}
L_f^2=(x+y)(1-x-y)k^2+x\Lambda _s^2+ym^2+(1-x-y)\sigma ^2,  \tag{41b}
\end{equation}
where $\eta \rightarrow 0^{+}$. It is seen that the Wick rotation can be
applied and the four-dimensional integral can be worked out immediately.
From Eqs. (22) and (41) we obtain 
\begin{equation}
\begin{array}{c}
a_s^f(k)=\frac{g_s^2}{16\pi ^2}\diint_0^\infty d\sigma ^2dm^2\dint_0^1dz~%
\frac{(\Lambda _s^2-m_s^2)}{(\Lambda _s^2-m^2)}~z~ \\ 
\qquad \times f_\alpha (-\sigma ^2)h_s(-m^2)\ln \frac{L_f^2(z,z,\sigma
^2,m^2,k^2)}{L_f^2(0,z,\sigma ^2,m^2,k^2)}
\end{array}
,  \tag{42a}
\end{equation}
\begin{equation}
\begin{array}{c}
b_s^f(k)=\frac{g_s^2}{16\pi ^2}\frac{M_t}M\diint_0^\infty d\sigma
^2dm^2\dint_0^1dz~\frac{(\Lambda _s^2-m_s^2)}{(\Lambda _s^2-m^2)} \\ 
\qquad \times f_\beta (-\sigma ^2)h_s(-m^2)\ln \frac{L_f^2(0,z,\sigma
^2,m^2,k^2)}{L_f^2(z,z,\sigma ^2,m^2,k^2)}
\end{array}
,  \tag{42b}
\end{equation}
\begin{equation}
L_f^2(x,z,\sigma ^2,m^2,k^2)=z(1-z)k^2+(1-z)\sigma ^2+zm^2+x(\Lambda
_s^2-m^2).  \tag{42c}
\end{equation}
According to Eq. (42c) $L_f^2(z,z,\sigma ^2,m^2,k^2)$ is independent of $m^2$
and in the neigbourhood of $m^2=\Lambda _s^2$ 
\begin{equation}
\ln \frac{L_f^2(0,z,\sigma ^2,m^2,k^2)}{L_f^2(z,z,\sigma ^2,m^2,k^2)}=\frac
z{L_f^2(z,z,\sigma ^2,m^2,k^2)}(m^2-\Lambda _s^2)+\cdots ,  \tag{43}
\end{equation}
thus the integrand is well behaved at $m^2=\Lambda _s^2$ even if $\Lambda
_s^2>th_s$. Similarly from Eq. (41) 
\begin{equation}
\begin{array}{c}
a_v^f(k)=\frac{g_v^2}{8\pi ^2}\diint_0^\infty d\sigma ^2dm^2\dint_0^1dz~%
\frac{(\Lambda _v^2-m_v^2)}{(\Lambda _v^2-m^2)}~z \\ 
~\qquad \times f_\alpha (-\sigma ^2)h_v(-m^2)\ln \frac{L_f^2(z,z,\sigma
^2,m^2,k^2)}{L_f^2(0,z,\sigma ^2,m^2,k^2)}
\end{array}
,  \tag{44a}
\end{equation}
\begin{equation}
\begin{array}{c}
b_v^f(k)=\frac{g_v^2}{4\pi ^2}\frac{M_t}M\diint_0^\infty d\sigma
^2dm^2\dint_0^1dz~\frac{(\Lambda _v^2-m_v^2)}{(\Lambda _v^2-m^2)} \\ 
\qquad \times f_\beta (-\sigma ^2)h_v(-m^2)\ln \frac{L_f^2(z,z,\sigma
^2,m^2,k^2)}{L_f^2(0,z,\sigma ^2,m^2,k^2)}
\end{array}
.  \tag{44b}
\end{equation}
Eqs. (42) and (44) combined with Eqs. (23, 26, and 28-30) build a closed
set, which has been solved by iteration. However, one notes that instead of
Eqs.(26b and c) one should now write 
\begin{equation}
\beta (k^2)=\frac M{\pi M_t}Im\left[ \frac{1+b_s(k^2)+b_v(k^2)}{D(k^2)}%
\right] ,  \tag{26d}
\end{equation}
\begin{equation}
D(k^2)=(1+a_s(k^2)+a_v(k^2))^2k^2+(1+b_s(k^2)+b_v(k^2))^2M^2,  \tag{26e}
\end{equation}
while between $M_t$ and $M$ we have 
\begin{equation}
(1+a_s(-M_t^2)+a_v(-M_t^2))^2M_t^2=(1+b_s(-M_t^2)+b_v(-M_t^2))^2M^2. 
\tag{27}
\end{equation}
The numerical results are represented graphically in Fig. 12 to 15 for $%
\Lambda _s=\Lambda _v=5.0676fm^{-1}$($1GeV$). The other parameters are the
same as chosen previously. According to our calculation no noticeable
difference between scheme BP and FSC exists, thus only the latter is
plotted. From the figures it is seen that the self-consistency is almost of
no effect. Comparing Figs. 12-13 with Figs. 6-7 one notes that the two set
of ($a,b$) differ significantly, especially with respect to their asymptotic
behavior. However, the results given in Figs. 8 (10) and 14 (15) for the
baryon (meson) spectral functions are qualitatively similar. In Table 5 we
have presented the values of $I_{b\text{ }}$ and $I_\lambda $ calculated
according to the method of renormalization and of form factors. If we adjust
the parameters, it is possible to make the two sets approach each other
better. This suggests that for the calculation of hadron propagators the two
methods can yield closely approximated results. We have found that all the
ghost poles are eliminated. The reason seems intimately related with the
change of the asymptotic behavior caused by the vertex correction. For the
baryon self-energy we see that as $|k^2|\rightarrow \infty ,$ the
Log-function $|\ln [K^2(-M_t^2)/K^2(k^2)]|\rightarrow \infty $ in Eqs. (24,
25), while in Eqs. (42, 44) we have $|\ln [L_f^2(z,z,\sigma
^2,m^2,k^2)/L_f^2(0,z,\sigma ^2,m^2,k^2)]|\rightarrow 0$. This fact also
explains the asymptotic behavior mentioned above. As shown in Eqs. (39, 40)
the meson self-energies $\Pi _\lambda ^f$ are obtained by simply multiplying
the corresponding expressions given in Section 2 by $F_\lambda (k^2)$, which
tends to $|k^2|^{-1}$ as $|k^2|\rightarrow \infty $. From Fig. 16 and Eqs.
(28, 29) one sees that $i\Pi _\lambda (k^2)$ bends the straight line $%
k^2+m_\lambda ^2$ to intersect the abscissa axis ($k^2$-axis) and thus $%
\Delta _\lambda (k)$ gets a ghost pole. However no such bending will be
induced by $i\Pi _\lambda ^f=F_\lambda (k^2)i\Pi _\lambda (k^2)$ as shown in
Fig. 17, because if $k^2$ becomes large, $i\Pi _\lambda ^f$ will be
negligible compared with $k^2$. It is interesting to note that there will be
no ghost poles if $i\Pi _\lambda (k^2)$ bends the straight line upward or
the curve $D_\lambda =D_\lambda (k^2)$ turns upward before it intersects the 
$k^2$-axis.

\section{. Concluding remarks}

Under the self-consistent HF approximation we have solved the coupled set of
DS equations for the renormalized hadron propagators in the $\sigma -\omega $
model. Since this set can be solved rigorously, it provides a convenient
means for the study of the effects of self-consistency. Our calculation
shows that in the case of zero-density scheme PEP gives a quite good
approximation to the FSC scheme and there is no need to require
self-consistency in the meson propagators. In the $\sigma -\omega $ case the
self-consistency almost has no effect on the baryon propagator, since there
is a cancellation between the effects caused by the two kinds of mesons,
whereas for the $\sigma $ case a FSC calculation is also unnecessary,
because the iteration procedure converges quickly if an appropriate initial
input is chosen. Fig. 3 shows that the self-consistency diminishes the
continuum part of the baryon spectral representation predicted by scheme P.
From Tab. 2 one observes that this part is by no means always unimportant.
However, there is generally no need to consider scheme BP, since if scheme P
does not yield a good enough result, one may go one step further to use
scheme PEP. For the study of nuclear matter and nuclear structure one has to
take the case of non-zero density into account. The question whether in the
latter case the above conclusions still hold remains to be studied. Using
the method of phenomenological form factors, we have also investigated the
significance of vertex corrections. Our calculation confirms the results of
Refs. [4, 6, 7, 9], namely the ghost poles in the hadron propagators can be
eliminated by vertex functions with damping asymptotic behavior. In Ref. [7]
it was found that the baryon spectral function $A_R(\kappa )$ can be
negative for some values of real $\kappa $ if only the coupling with $\omega 
$-mesons is considered. In the latter case we have also found that $\alpha
(k^2)$ may be negative in a region close to the threshold. Though, if in
addition to $\omega $ other mesons are taken into account, the above problem
does not occur, it still has to be studied and is being under study, because
the coupling with $\omega $-mesons is of importance and according to their
definitions and physical meanings, both $A_R(\kappa )$ and $\alpha (k^2)$
should be $\geq 0$.

The work is supported in part by the National Natural Science Foundation of
China and the Foundation of Chinese Education Ministry.

\newpage\

\break 

\begin{center}
\textbf{Table}
\end{center}

\begin{description}
\item[Table 1]  : Different calculation schemes.
\end{description}

\begin{tabular}{|c|c|c|}
\hline
Name & $\Sigma _s$ ($\Sigma _v$) & $\Pi _s$ ($\Pi _v$) \\ \hline
P & $G_\Sigma ^0$, $\Delta _s^0$ ($\Delta _v^0$) & $G_\Sigma ^0$ \\ \hline
PEP & $G(P)$, $\Delta _s^0$ ($\Delta _v^0$) & $G(P)$ \\ \hline
EP & $G(P)$, $\Delta _s^0(P)$ ($\Delta _v^0(P)$) & $G(P)$ \\ \hline
BP & $G(BP)$, $\Delta _s^0$ ($\Delta _v^0$) & $G(BP)$ \\ \hline
FSC & $G(FSC)$, $\Delta _s^0(FSC)$ ($\Delta _v^0(FSC)$) & $G(FSC)$ \\ \hline
\end{tabular}

\begin{description}
\item[Table 2]  : Values of the baryon impurity factor $I_b$.
\end{description}

\begin{tabular}{|l|r|r|r|r|r|}
\hline
Scheme & P & PEP & EP & BP & FSC \\ \hline
$\overline{g}_s^2=0.5263$ & $0.4157$ & $0.3528$ & $0.3504$ & $0.3672$ & $%
0.3664$ \\ \hline
$\overline{g}_s^2=0.6517$ & $0.4638$ & $0.3253$ & $0.3227$ & $0.3756$ & $%
0.3752$ \\ \hline
\end{tabular}

\begin{description}
\item[Table 3]  : Values of the $\sigma $-meson impurity factor $I_\sigma $.

\begin{tabular}{|c|c|c|c|}
\hline
Scheme & P & PEP & FSC \\ \hline
$\overline{g}_s^2=0.5263$ & $0.1725$ & $0.1178$ & $0.1241$ \\ \hline
$\overline{g}_s^2=0.6517$ & $0.1557$ & $0.0931$ & $0.1017$ \\ \hline
\end{tabular}
\end{description}

\begin{description}
\item[Table 4]  : Ghost poles of the baryon and $\sigma $-meson propagators.
\end{description}

\begin{tabular}{|c|c|c|}
\hline
Propagator & Baryon & $\sigma $-meson \\ \hline
$\overline{g}_s^2=0.5263$ & $35.5356\pm i253.5311$ & $246.7354$ \\ \hline
$\overline{g}_s^2=0.6517$ & $-13.6188\pm i127.7365$ & $177.4338$ \\ \hline
\end{tabular}

\begin{description}
\item[Table 5]  : Impurity factors calculated according to the method of
renormalization (I) and of form factors (II).

\begin{tabular}{|cccc|}
\hline
\multicolumn{1}{|c|}{Impurity factor} & \multicolumn{1}{c|}{$I_b$} & 
\multicolumn{1}{c|}{$I_\sigma $} & $I_\omega $ \\ \hline
\multicolumn{1}{|c|}{I} & \multicolumn{1}{c|}{$0.2905$} & 
\multicolumn{1}{c|}{$0.1334$} & $0.1870$ \\ \hline
\multicolumn{1}{|c|}{II} & \multicolumn{1}{c|}{$0.2701$} & 
\multicolumn{1}{c|}{$0.0459$} & $0.0643$ \\ \hline
\end{tabular}
\end{description}

Table 1 on P$_9$, Tables 2-4 on P$_{10}$, Table 5 on P$_{14}$.

\break 

\begin{center}
\textbf{Figure captions}
\end{center}

\begin{description}
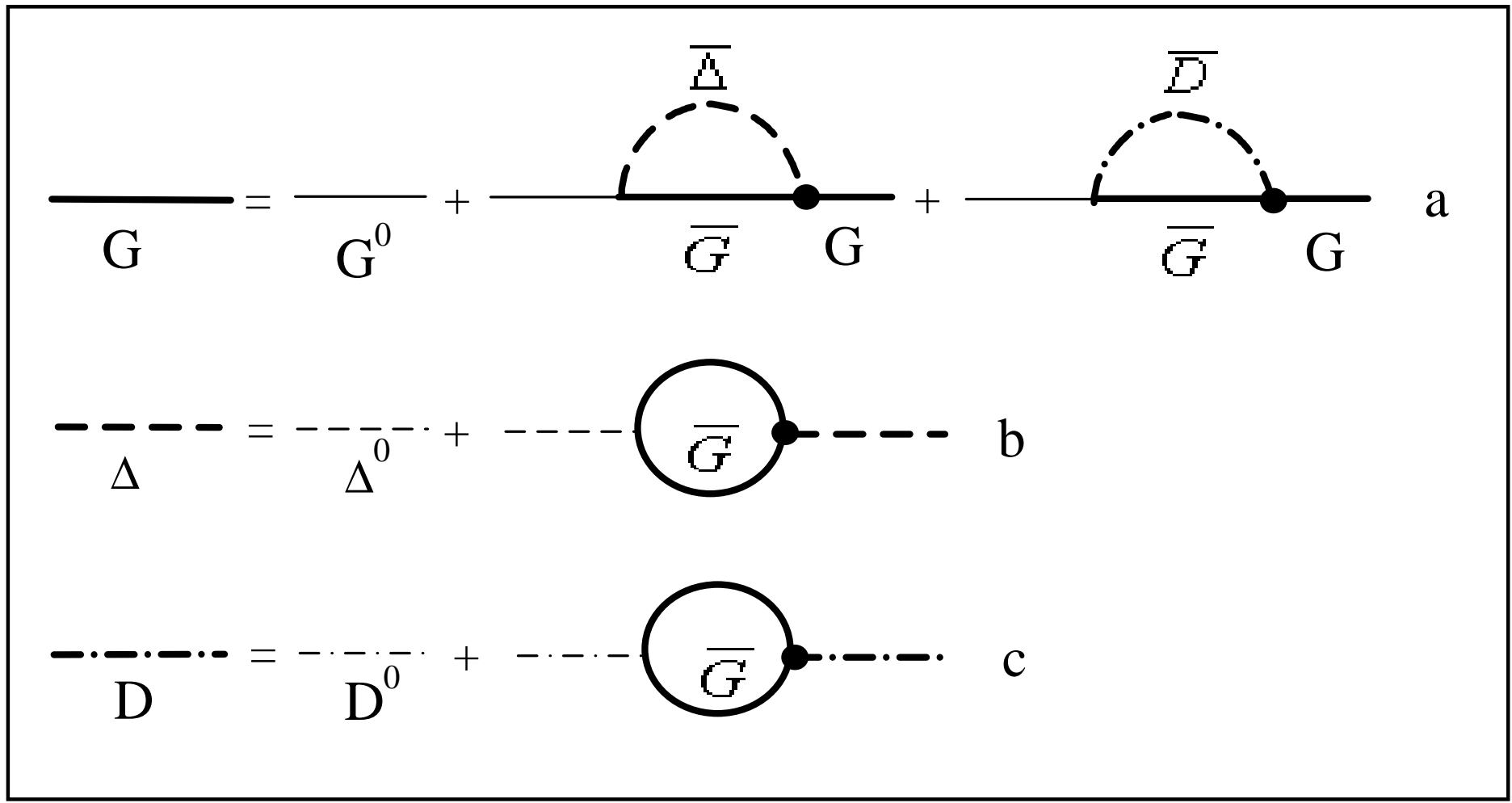
\item[Fig. 1]  : Diagrammatic representation of the different
self-consistent (dressed) HF schemes. a. the baryon propagator, b. the $%
\sigma $-meson propagator, c. the $\omega $-meson propagator.

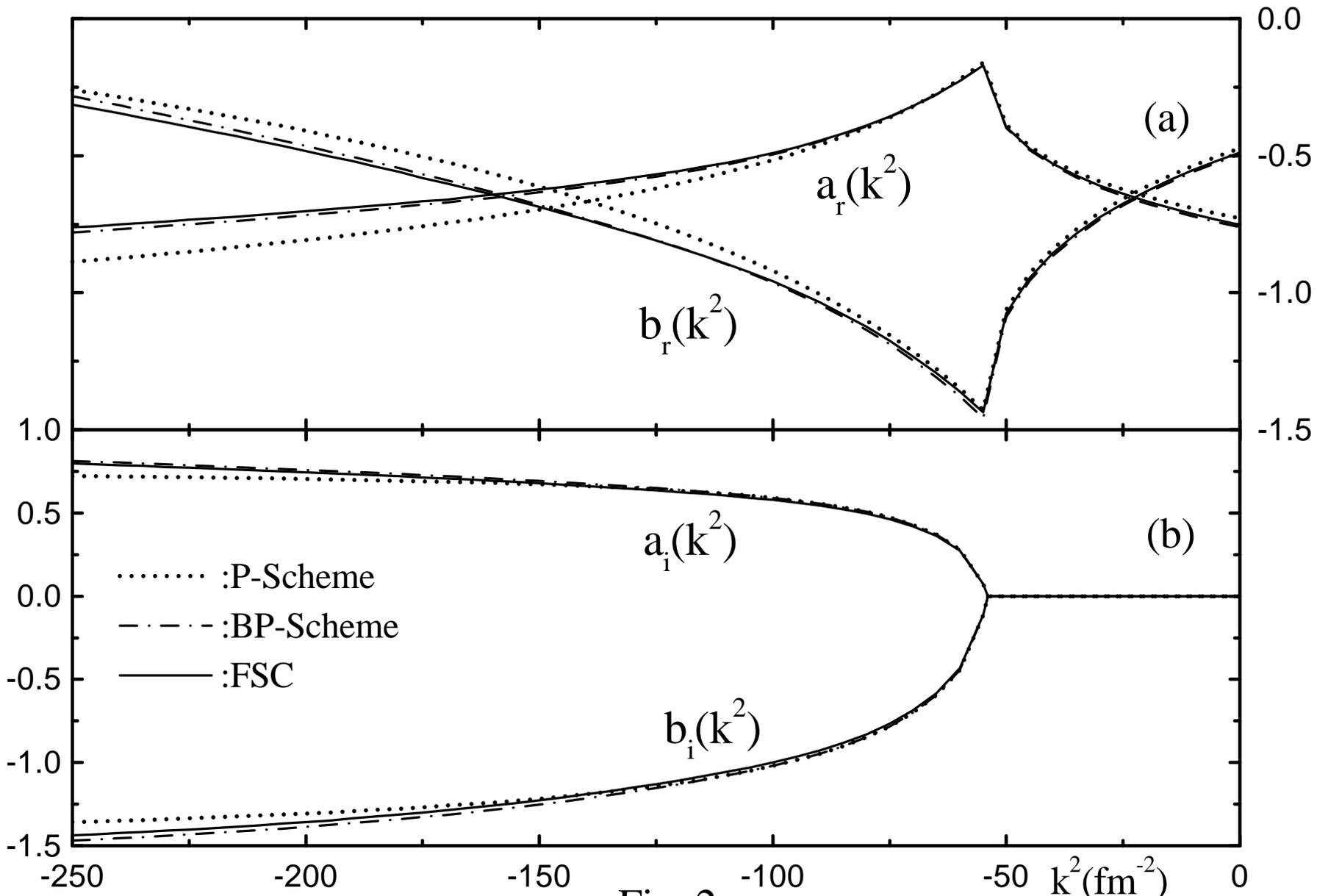
\item[Fig. 2]  : Numerical results of ($a_s$, $b_s$). (a) the real part, (b)
the imaginary part. [$BP\simeq FSC$, $\simeq $ means indistinguishable or
almost indistinguishable.]
\end{description}

\begin{description}
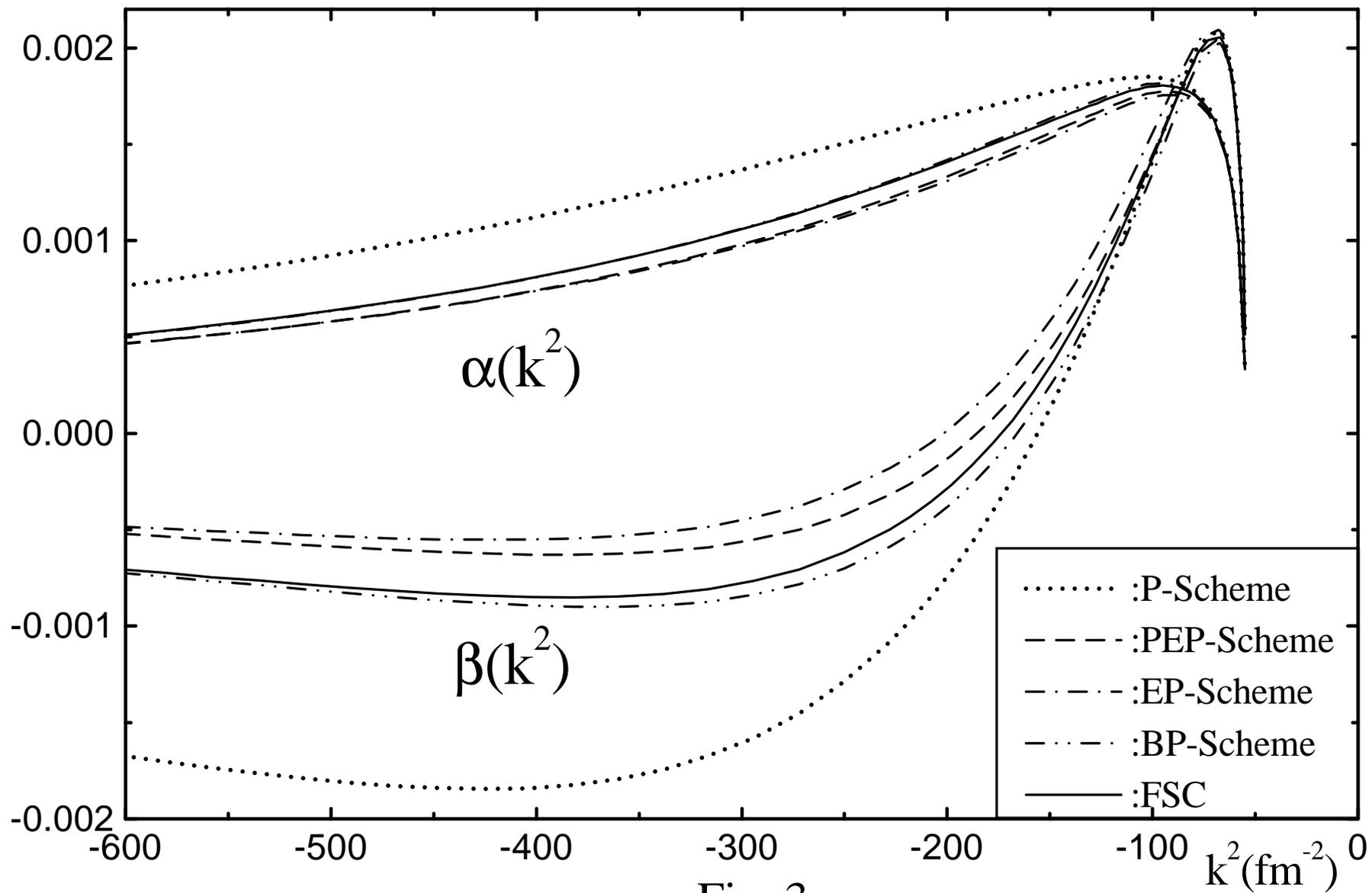
\item[Fig. 3]  : The baryon spectral functions $\alpha (k^2)$ and $\beta
(k^2)$. [$\alpha (BP)\simeq \alpha (FSC)$; $\alpha (PEP)\simeq \alpha (EP)$]

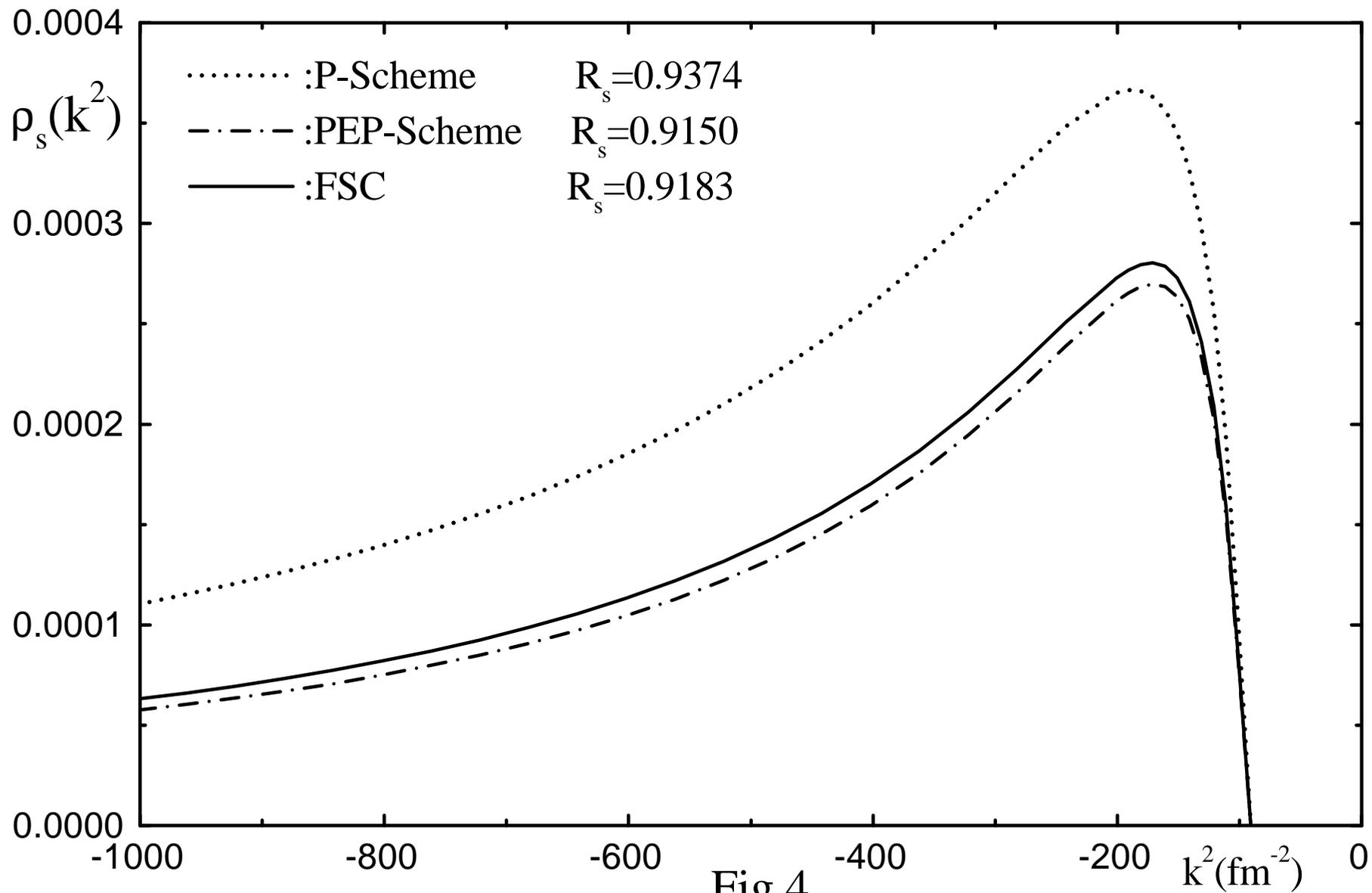
\item[Fig. 4]  : The $\sigma $-meson spectral functions $\rho _s(k^2)$.

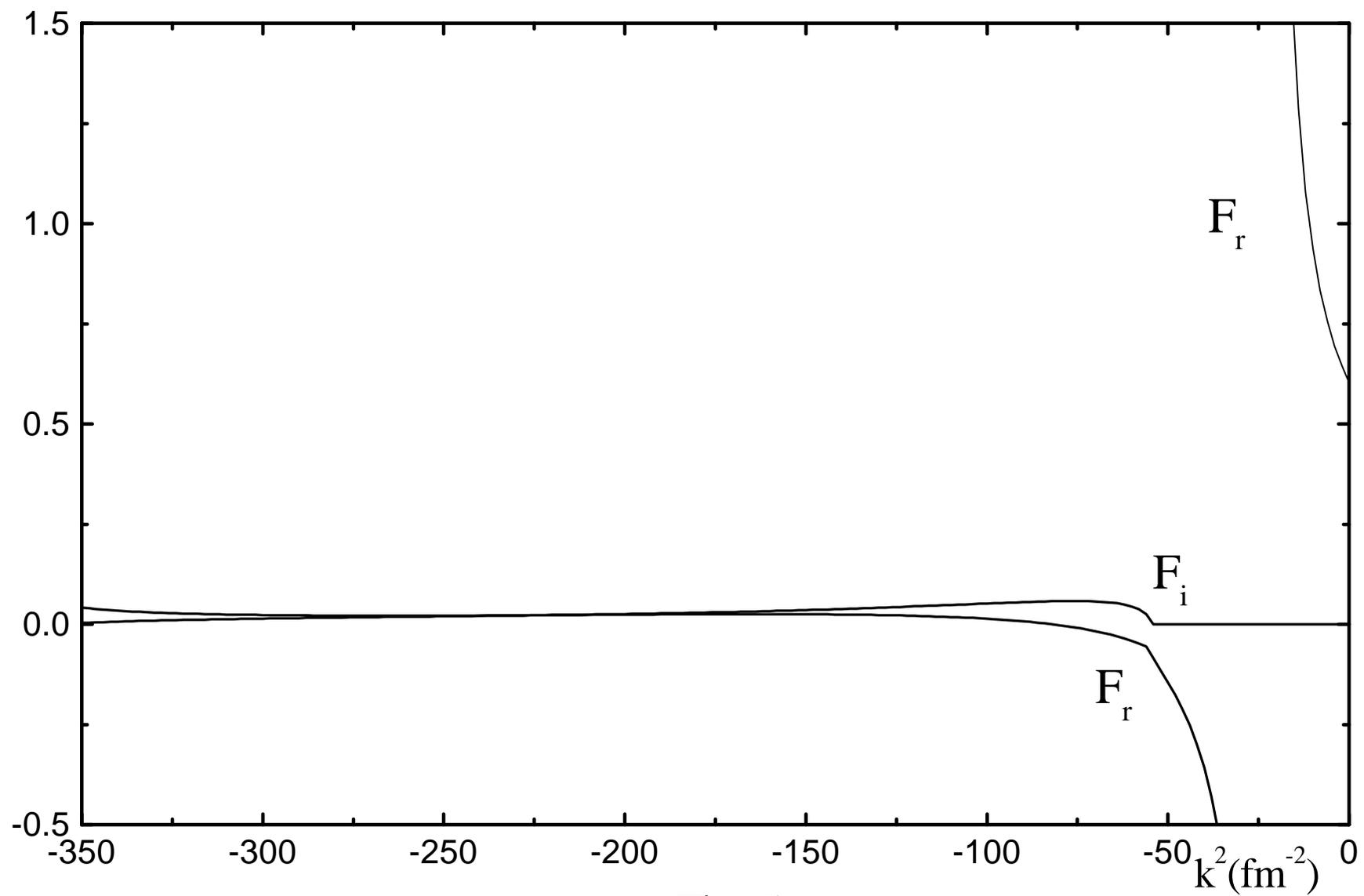
\item[Fig. 5]  : Graphical representation for $F(+)=<iG(k)>_{+}$,where $%
F_r=Re<iG_{HF}(FSC)>_{+}$, $F_r=Im<iG_{HF}(FSC)>_{+}$.

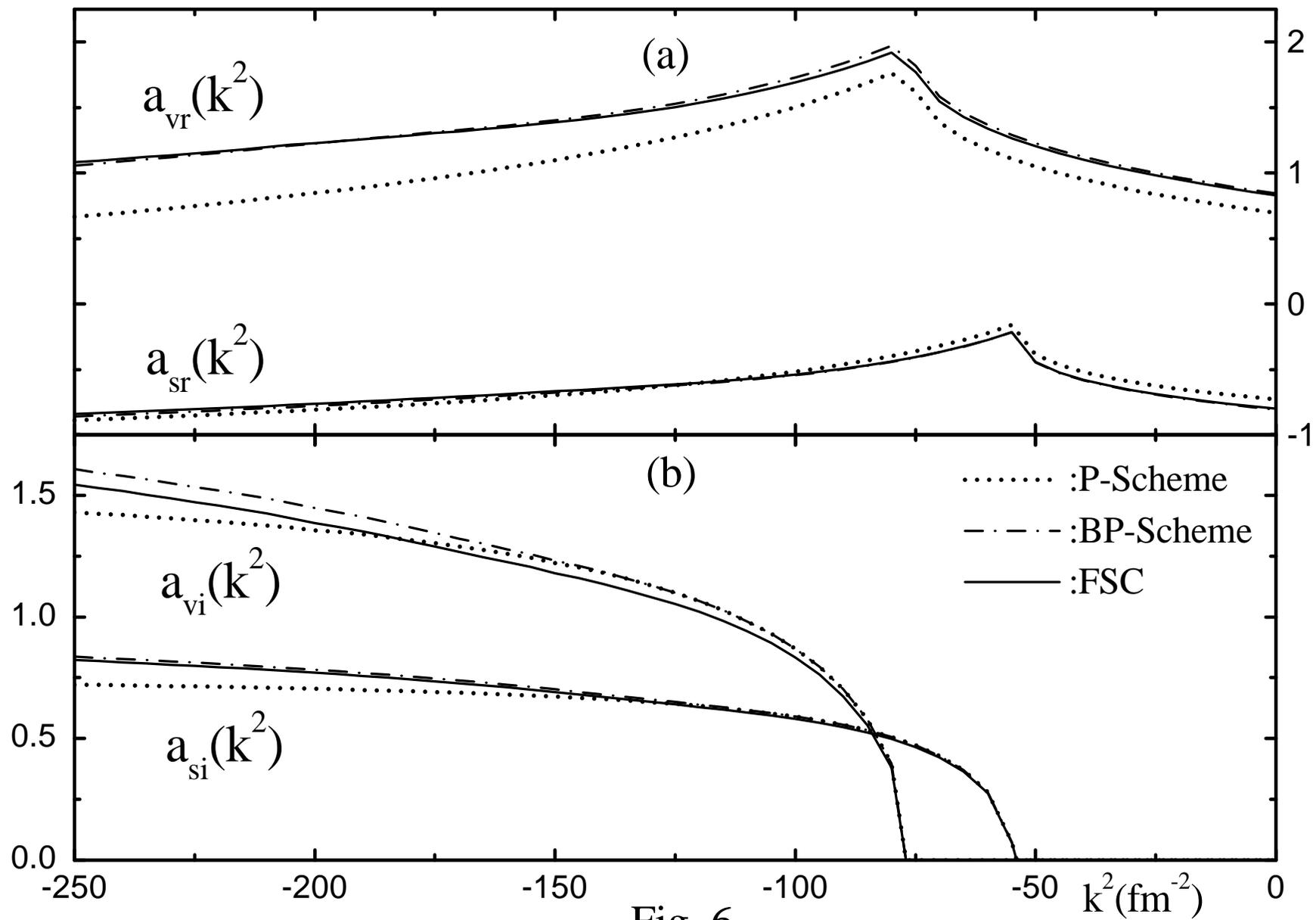
\item[Fig. 6]  : Numerical results of $a_s$ and $a_v$. (a) the real part,
(b) the imaginary part. [$a_{sr}(BP)\simeq a_{sr}(FSC)$; $a_{si}(BP)\simeq
a_{si}(FSC)$; $a_{vr}(BP)\simeq a_{vr}(FSC)$]

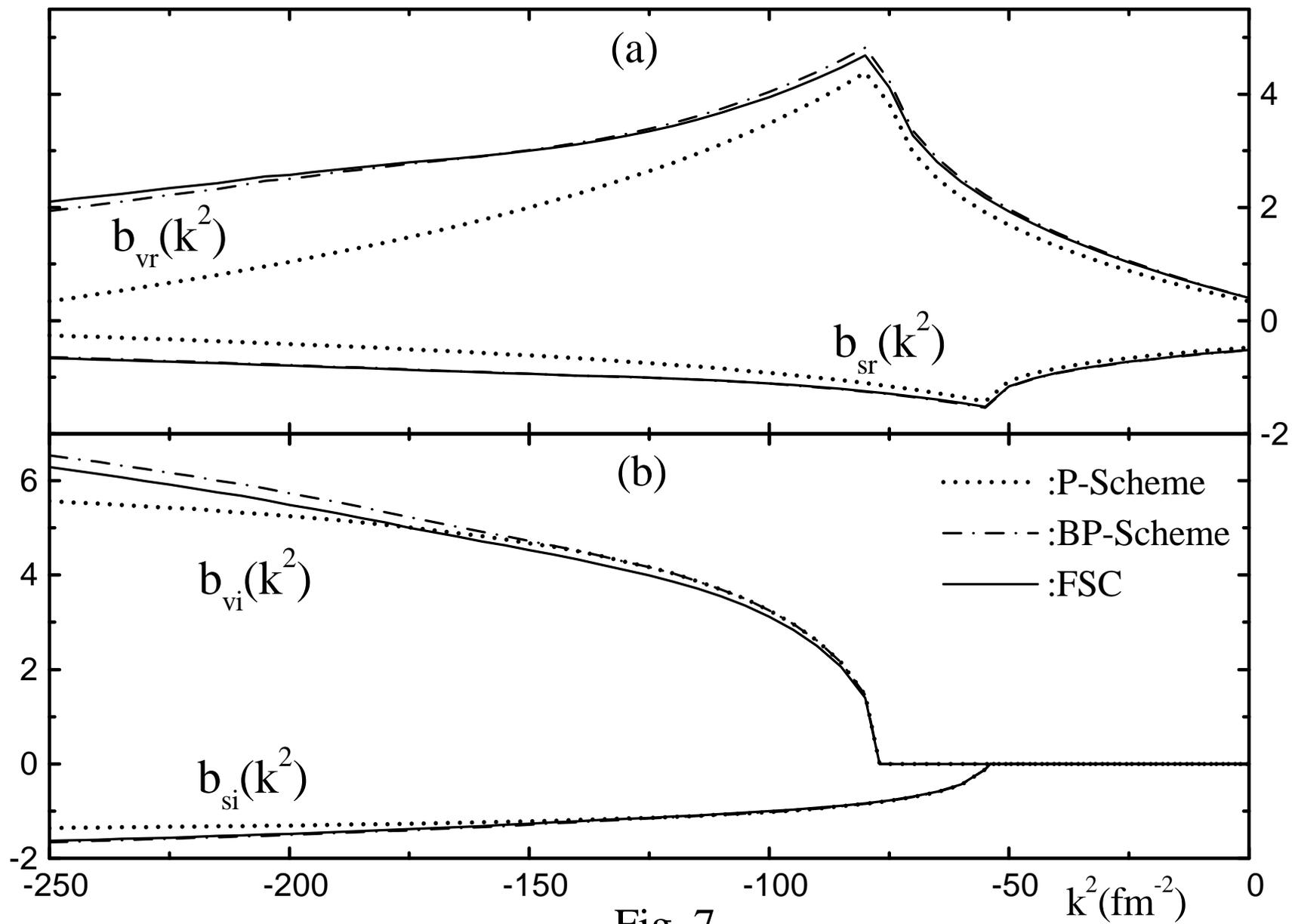
\item[Fig. 7]  : Numerical results of $b_s$ and $b_v$. (a) the real part,
(b) the imaginary part. [$b_{sr}(BP)\simeq b_{sr}(FSC)$; $b_{si}(BP)\simeq
b_{si}(FSC)$; $b_{vr}(BP)\simeq b_{vr}(FSC)$]

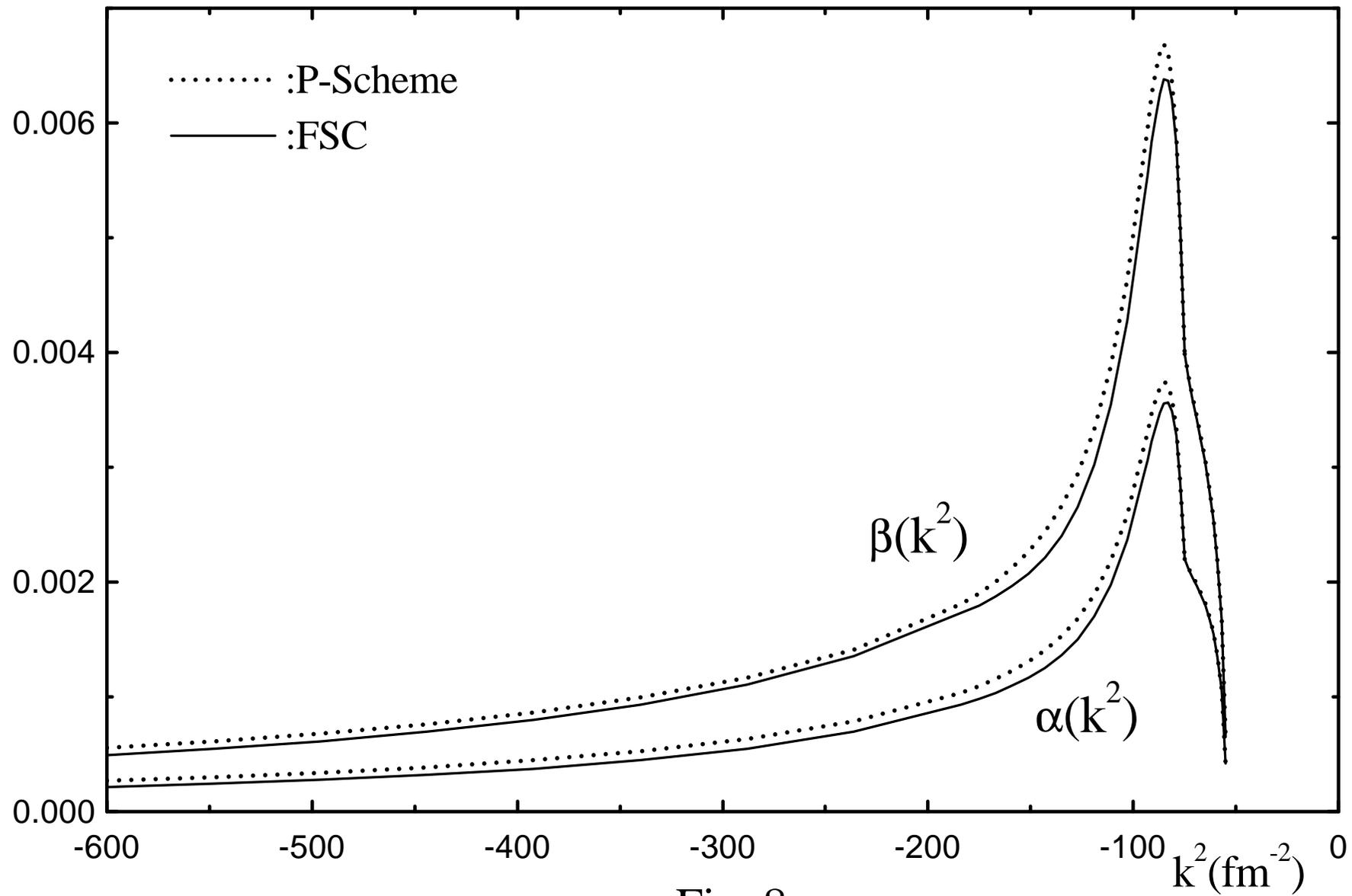
\item[Fig. 8]  : The baryon spectral functions $\alpha (k^2)$ and $\beta
(k^2)$ calculated with baryon coupling with both $\sigma $ and $\omega $
meson.

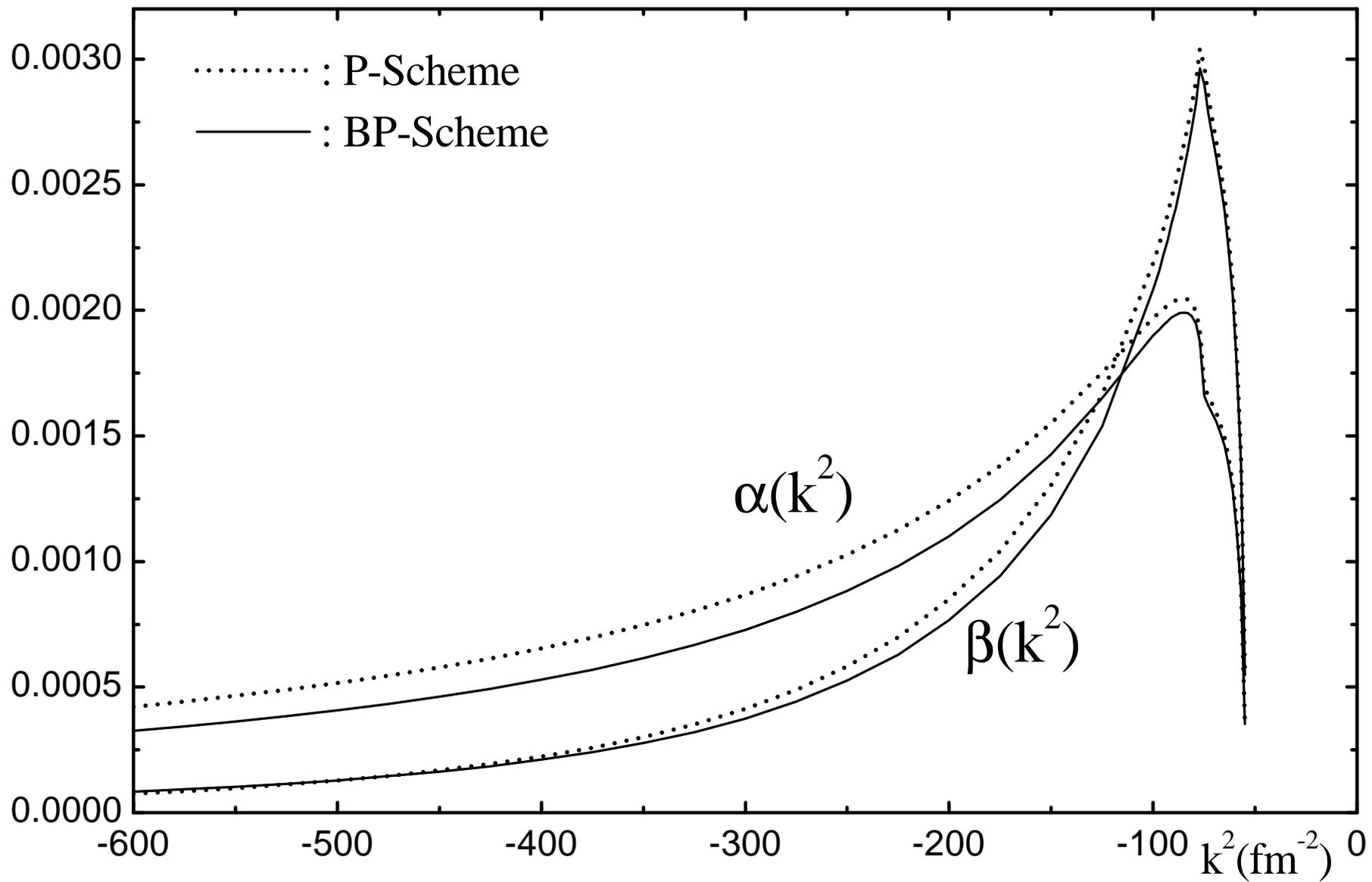
\item[Fig. 9]  : The baryon spectral functions $\alpha (k^2)$ and $\beta
(k^2)$ for $\bar g_v^2=0.34$.

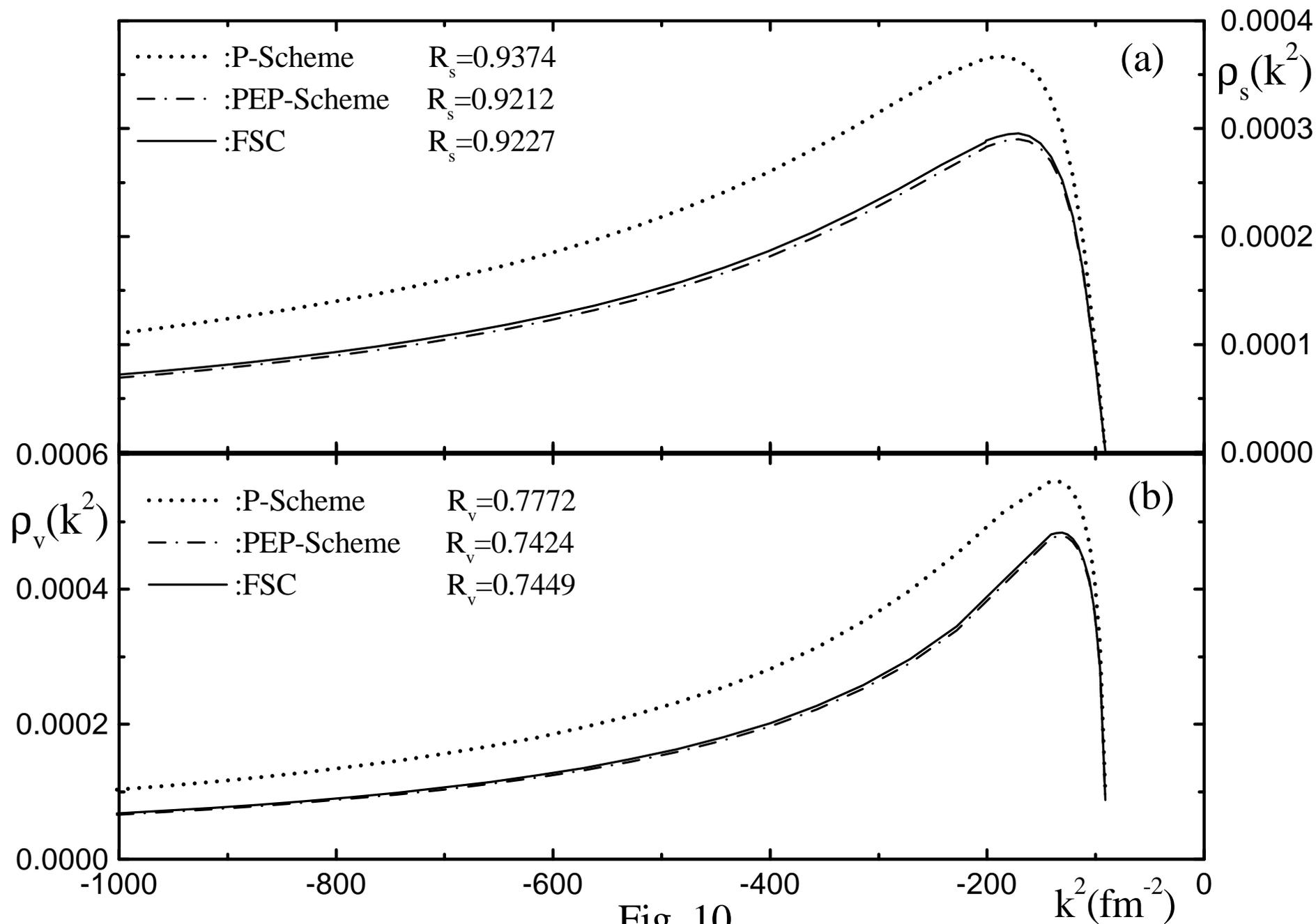
\item[Fig. 10]  : The meson spectral functions $\rho _\lambda (k^2)$. (a) $%
\sigma $-meson: $\lambda =\sigma $; (b) $\omega $-meson: $\lambda =\omega $.
[$PEP\simeq FSC$]

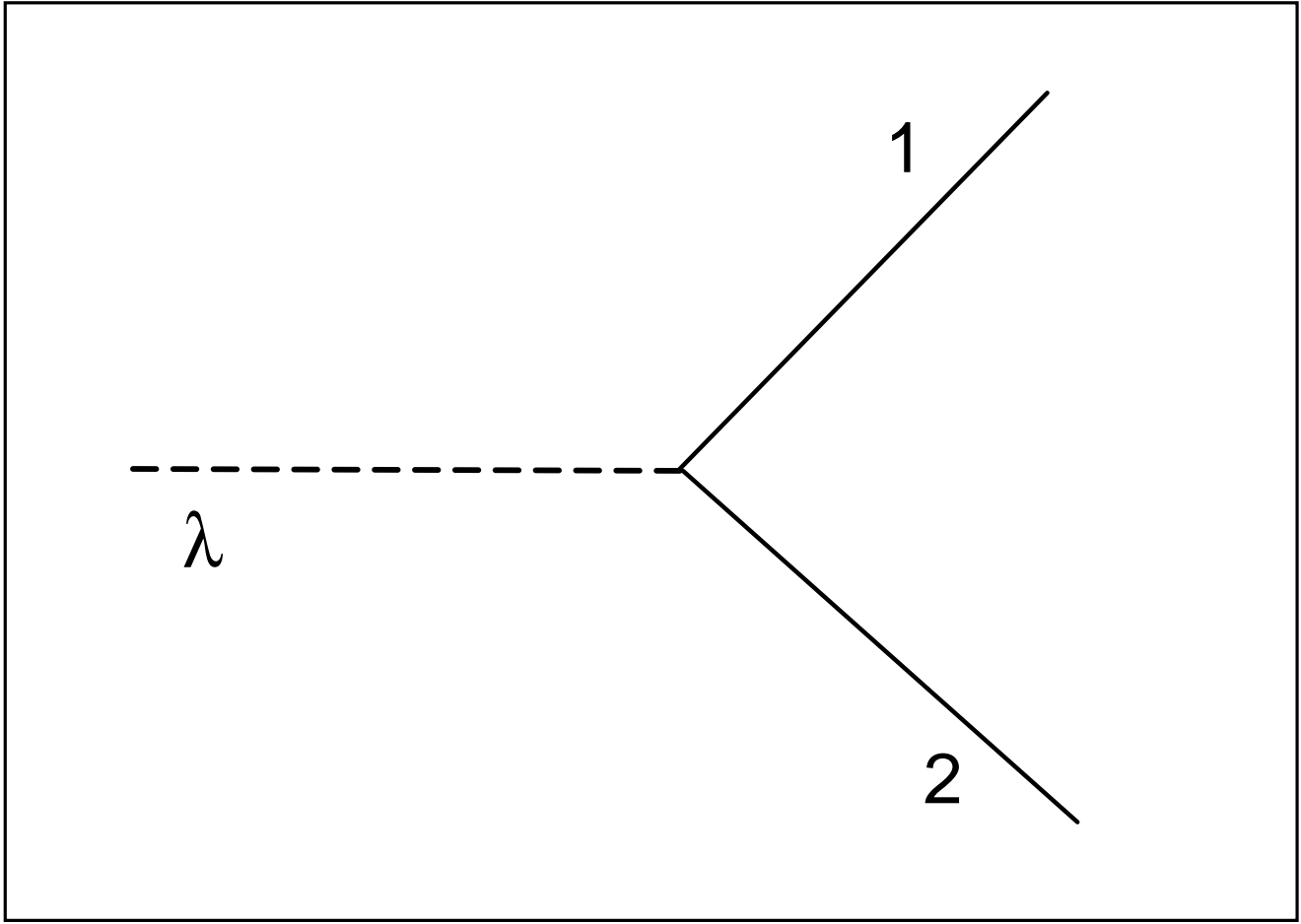
\item[Fig. 11]  : The baryon-meson vertices.

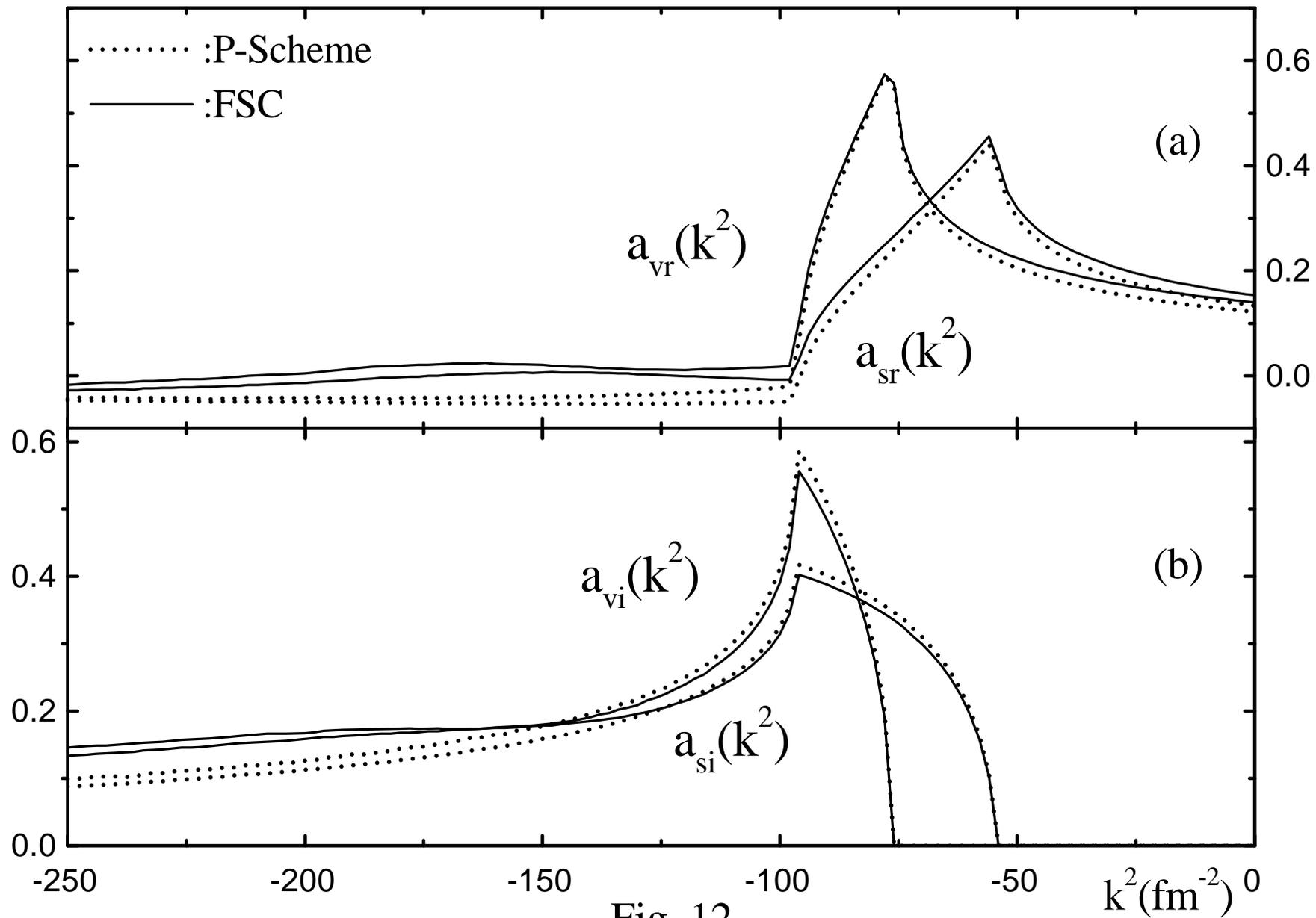
\item[Fig. 12]  : Numerical results of ($a_s$, $a_v$) [FF]. where the square
bracket [FF] indicates that the results are obtained by the method of form
factors.

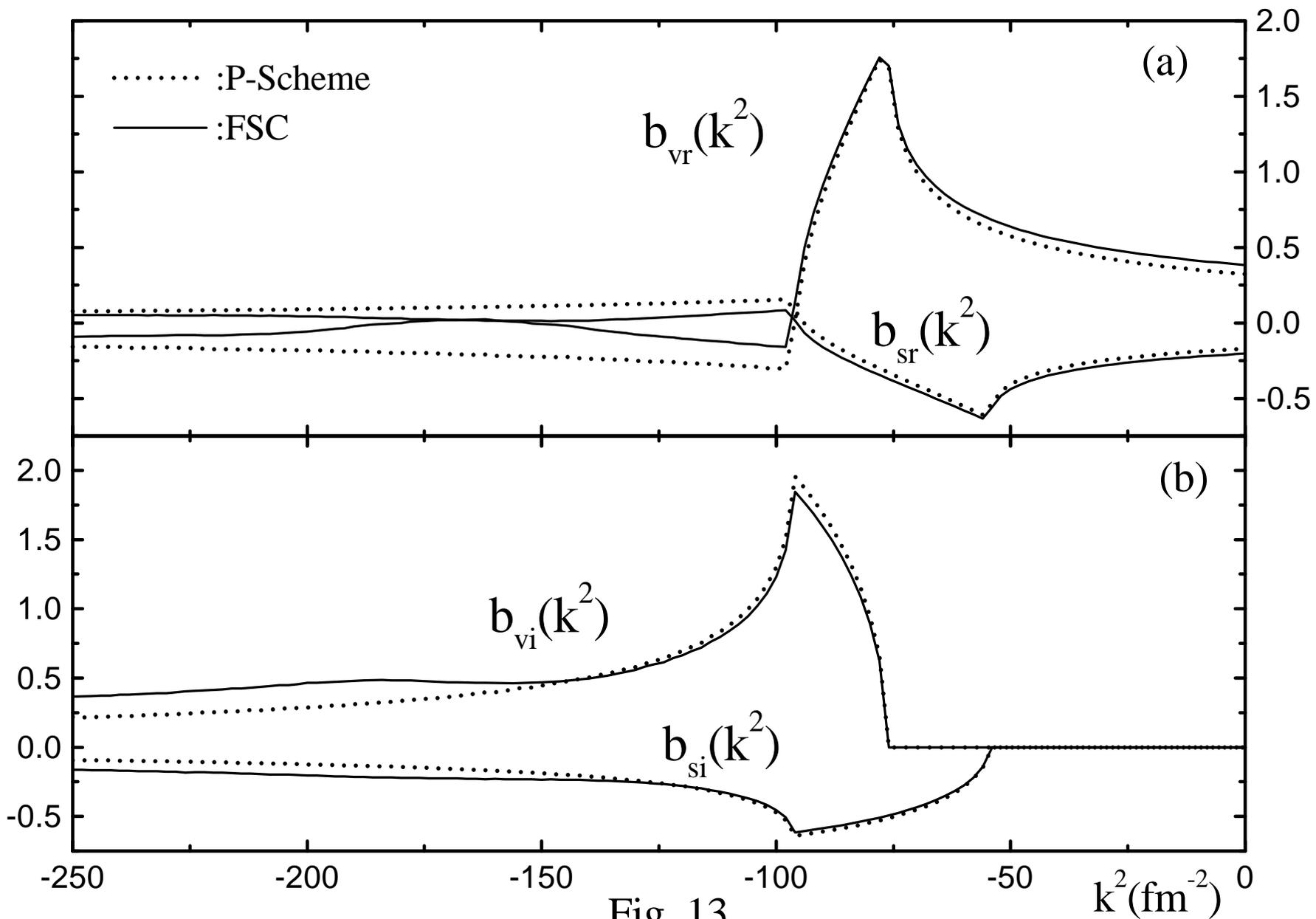
\item[Fig. 13]  : Numerical results of ($b_s$, $b_v$) [FF].

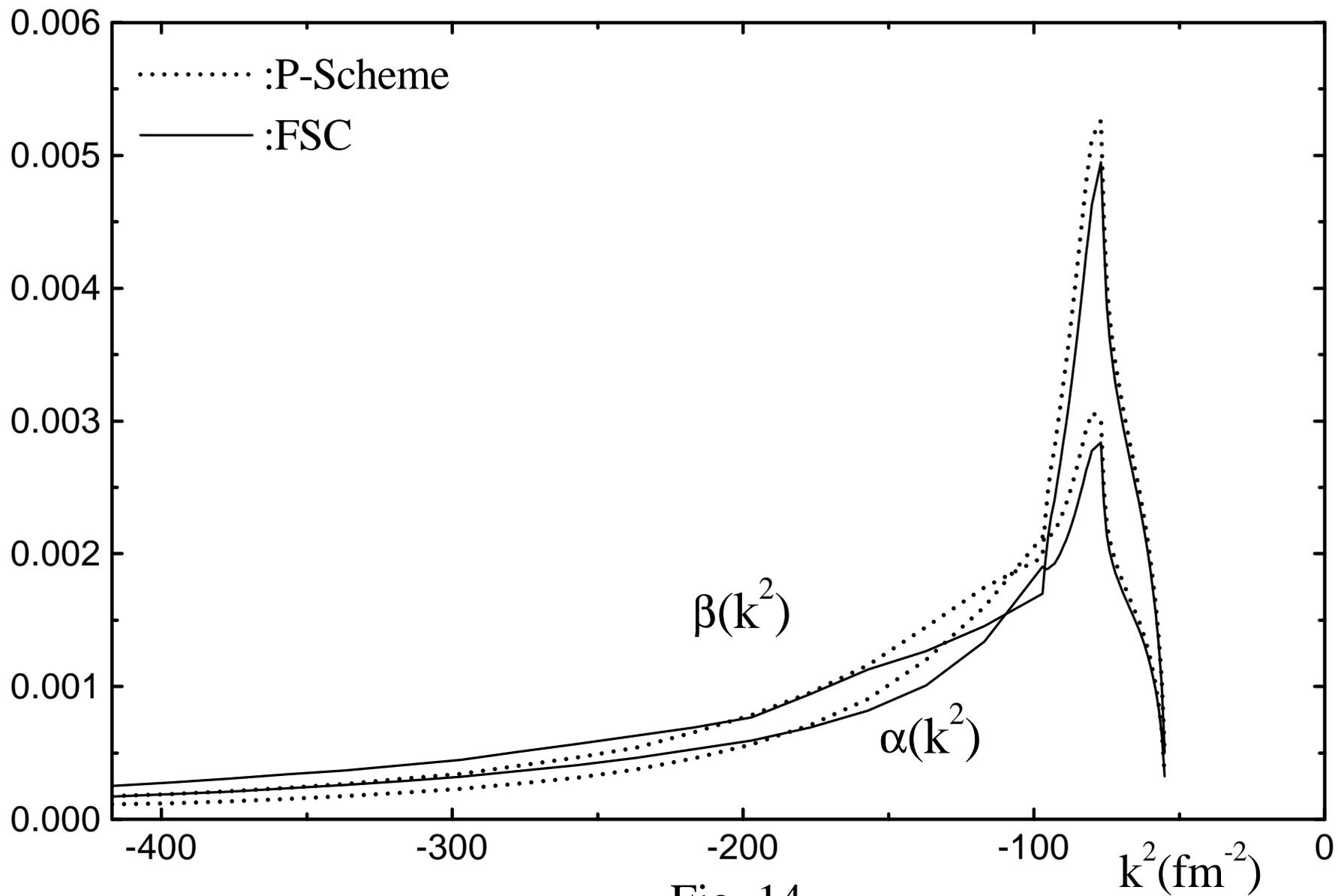
\item[Fig. 14]  : The baryon spectral functions $\alpha (k^2)$ and $\beta
(k^2)$.

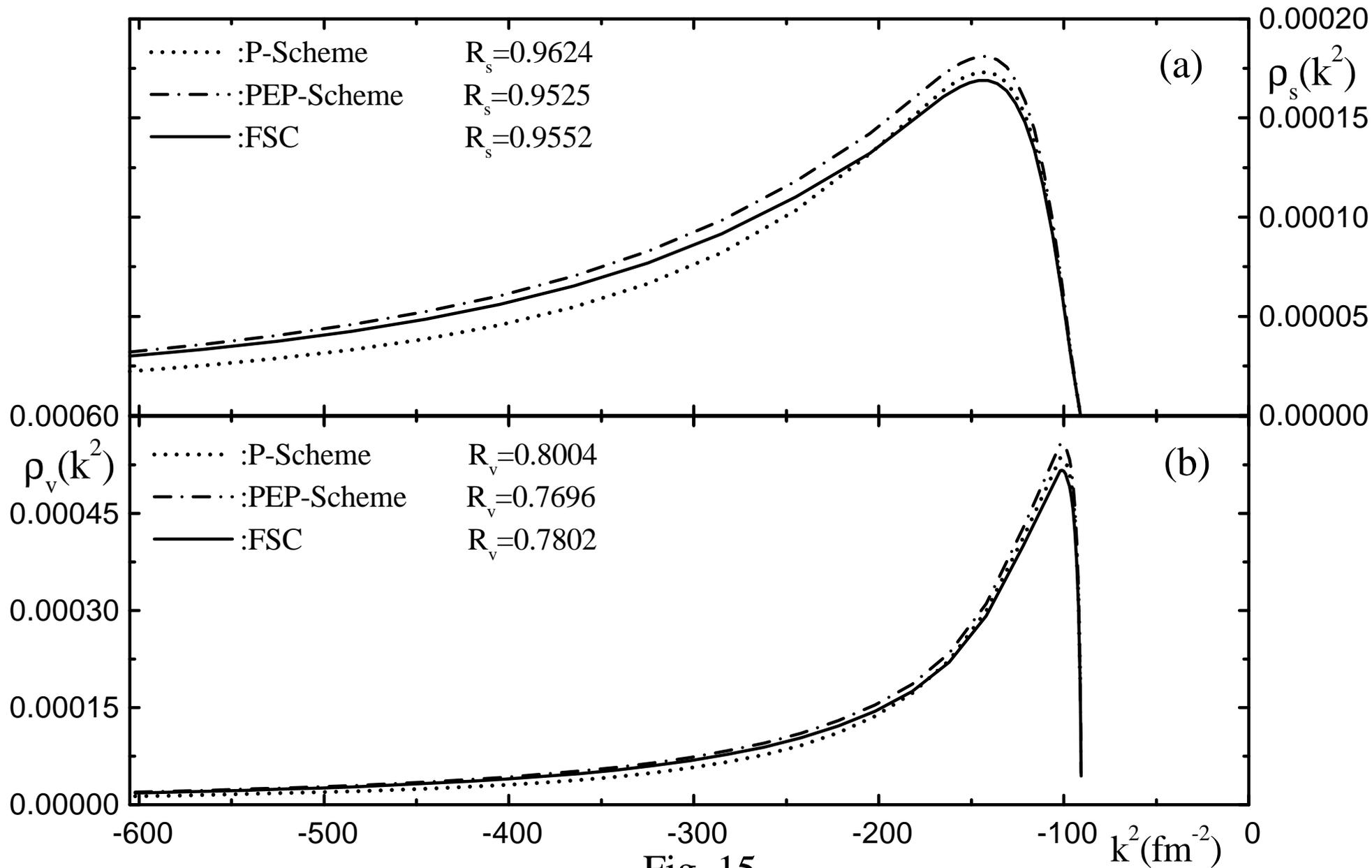
\item[Fig. 15]  : The meson spectral functions $\rho _\lambda (k^2)$ [FF].
(a) $\sigma $-meson: $\lambda =s$; (b) $\omega $-meson: $\lambda =\omega $.

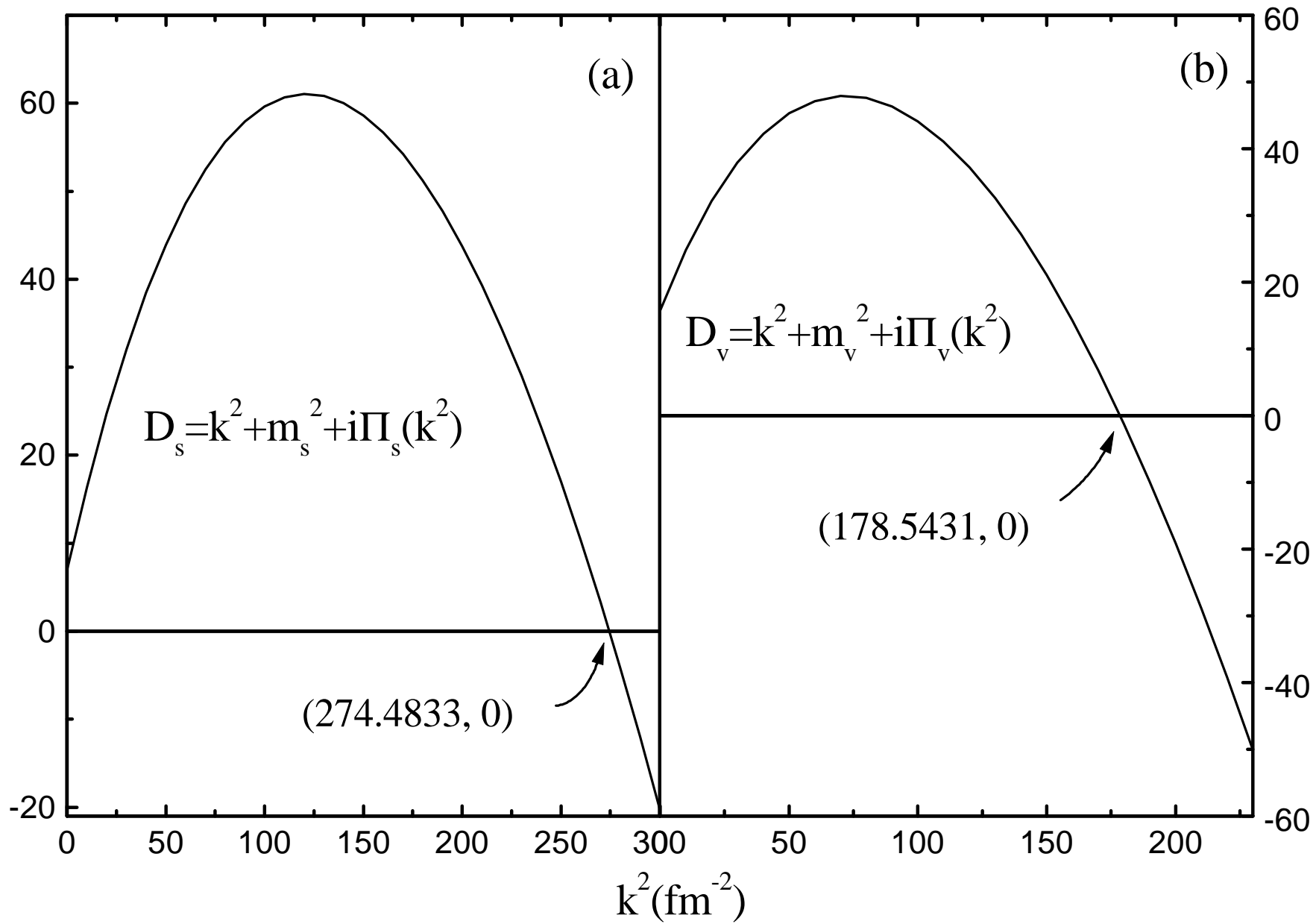
\item[Fig. 16]  : Plot of $D_\lambda =D_\lambda (k^2)$ versus $k^2$. (a) $%
\sigma $-meson: $\lambda =s$; (b) $\omega $-meson: $\lambda =\omega $.

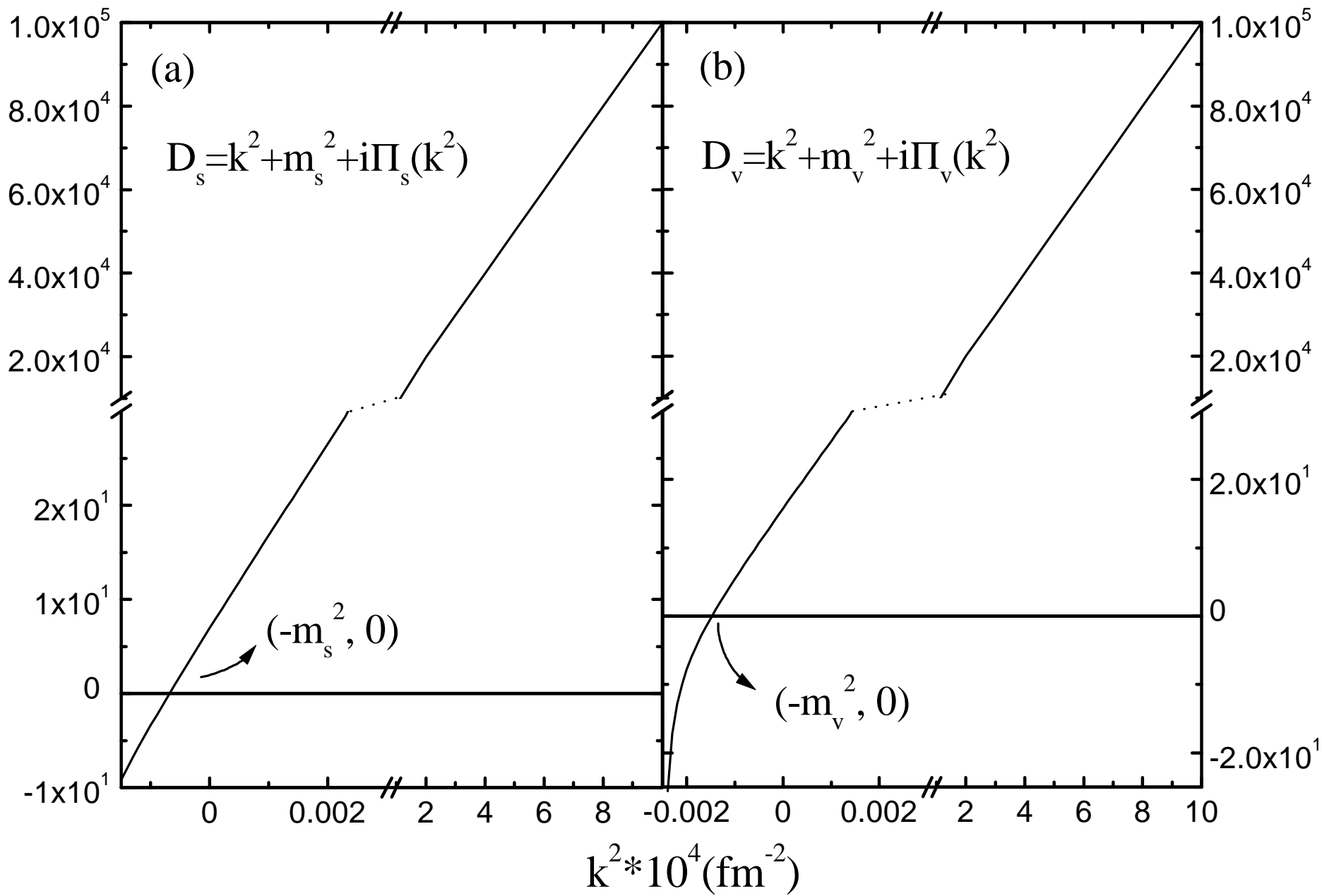
\item[Fig. 17]  : The same as Fig. 16 but calculated by the method of form
factors.
\end{description}

Fig. 1 on P$_2$, Figs. 2-4 on P$_9$, Figs. 5-7 on P$_{10}$, Figs. 8-10 on P$%
_{11}$, Figs. 12-17 on P$_{14}$.


\begin{thebibliography}{99}
\bibitem{}  W. D. Brown, R. D. Puff and L. Wilets, Phys. Rev. C 2 (1970) 331.

\bibitem{}  A. F. Bielajew and B. D. Serot, Ann. Phys. 156 (1984) 215; A. F.
Bielajew, Nucl. Phys. A 404 (1983) 428.

\bibitem{}  B. D. Serot and J. D. Walecka, Adv. Nucl. Phys. 16 (1986) 1; B.
D. Serot, Rep. Prog. Phys. 55 (1992) 1855.

\bibitem{}  M. E. Bracco, A. Eiras, G. Krein and L. Wilets, Phys. Rev. C 49
(1994) 1299.

\bibitem{}  J. Milana, Phys. Rev. C 44 (1991) 527.

\bibitem{}  M. P.\ Allendes and B. D. Serot, Phys. Rev. C 45 (1992) 2975.

\bibitem{}  G. Krein, M. Nielsen, R. D. Puff and L. Wilets, Phys. Rev. C 47
(1993) 2485.

\bibitem{}  S. S. Wu, J. M. Zhu, K. Z. Liu and Y. J. Yao, Eur. Phys. J. A 6
(1999) 345; Y. J. Yao, H. X. Zhang, J. M. Zhu, K. Z. Liu and S. S. Wu, Chin.
Phys. Lett. 17 (2000) 720.

\bibitem{}  C. A. da Rocha, G. Krein and L. Wilets, Nucl. Phys. A 616 (1997)
625.

\bibitem{}  N. M. Kroll, T. D. Lee and B. Zumino, Phys. Rev. 157 (1967) 1376.

\bibitem{}  D. Lurie, Particles and fields ( Interscience, New York, 1968).

\bibitem{}  S. S. Wu, Y. J. Yao, Eur. Phys. J. A 3 (1998) 49.

\bibitem{}  J. D. Bjorken and S. D. Drell, Relativistic Quantum Fields
(McGraw-Hill, New York, 1965).

\bibitem{}  B. C. Pearce and B. K. Jennings, Nucl. Phys. A 528 (1991) 655;
F. Gross, J. W. Van Orden and K. Holinde, Phys. Rev. C 45 (1992) 2094.
\end{thebibliography}
\end{document}